\documentclass{aa}  
\usepackage{graphicx}
\usepackage{txfonts}
\usepackage{threeparttable}
\usepackage{amsmath}    
\usepackage{amssymb}    
\usepackage[left = `, right = ', leftsub = ``, rightsub = '']{dirtytalk}
\usepackage{multirow}
\usepackage{gensymb}
\usepackage{color}
\usepackage[singlelinecheck=off]{subcaption}

\usepackage{hyperref}

\usepackage[dvipsnames]{xcolor}
\definecolor{myblue}{HTML}{1F77B4}
\definecolor{mygreen}{HTML}{2CA02C}
\definecolor{myred}{HTML}{D62728}
\definecolor{mymagenta}{HTML}{D33682}
\definecolor{codepurple}{HTML}{C42043}

\hypersetup{
bookmarks=true,         
unicode=true,           
colorlinks=true,        
linkcolor=myblue,        
citecolor=myblue,       
urlcolor=myblue        
}

\begin{document}

   \title{Polarimetry of Hydrogen-Poor Superluminous Supernovae}

  \author{ M. Pursiainen        \inst{1} \and
            G. Leloudas         \inst{1} \and
            A. Cikota           \inst{2} \and
            M. Bulla            \inst{3,4,5} \and
            C. Inserra          \inst{6} \and
            F. Patat            \inst{7} \and
            J. C. Wheeler       \inst{8} \and
            A. Aamer            \inst{9} \and
            A. Gal-Yam          \inst{10} \and
            J, Maund            \inst{11} \and
            M. Nicholl          \inst{9} \and
            S. Schulze          \inst{12} \and
            J. Sollerman        \inst{12} \and
            Y. Yang             \inst{13}
            }
%

   \institute{DTU Space, National Space Institute, Technical University of Denmark, Elektrovej 327, 2800 Kgs. Lyngby, Denmark\\
              \email{miipu@space.dtu.dk}
        \and
            Gemini Observatory / NSF's NOIRLab, Casilla 603, La Serena, Chile
        \and
            Department of Physics and Earth Science, University of Ferrara, via Saragat 1, I-44122 Ferrara, Italy
        \and
            INFN, Sezione di Ferrara, via Saragat 1, I-44122 Ferrara, Italy
        \and
            INAF, Osservatorio Astronomico d’Abruzzo, via Mentore Maggini snc, 64100 Teramo, Italy
        \and
            Cardiff Hub for Astrophysics Research and Technology, School of Physics \& Astronomy, Cardiff University, Queens Buildings, The Parade, Cardiff, CF24 3AA, UK 
        \and
            European Southern Observatory, Karl-Schwarzschild-Str. 2, D-85748 Garching b. M\"unchen, Germany
        \and
            Department of Astronomy, University of Texas at Austin, Austin, TX, USA
        \and
            Birmingham Institute for Gravitational Wave Astronomy and School of Physics and Astronomy, University of Birmingham, Birmingham, B15 2TT, UK
        \and
            Department of Particle Physics and Astrophysics, Weizmann Institute of Science, 76100, Rehovot, Israel
        \and
            Department of Physics and Astronomy, University of Sheffield, Hicks Building, Hounsfield Road, Sheffield, S3 7RH, UK
        \and
            The Oskar Klein Centre, Department of Astronomy, Stockholm University, AlbaNova, SE-10691 Stockholm, Sweden
        \and
            Department of Astronomy, University of California, Berkeley, CA 94720-3411, USA
}

   \date{}

 
  \abstract{We present linear polarimetry for seven hydrogen-poor superluminous supernovae (SLSNe-I) of which only one has previously published polarimetric data. The best-studied event is SN\,2017gci, for which we present two epochs of spectropolarimetry at $+3$\,d and $+29\,$d post-peak in rest frame, accompanied by four epochs of imaging polarimetry up to $+108$\,d. The spectropolarimetry at $+3$\,d shows increasing polarisation degree $P$ towards the redder wavelengths and exhibits signs of axial symmetry, but at $+29$\,d $P\sim0$ throughout the spectrum implying that the photosphere of SN\,2017gci evolved from a slightly aspherical configuration to a more spherical one in the first month post-peak. However, an increase of $P$ to $\sim0.5\%$ at $\sim+55$\,d accompanied by a different orientation of the axial symmetry compared to $+3$\,d implies the presence of additional sources of polarisation at this phase. The increase in polarisation is possibly caused by interaction with circumstellar matter (CSM) as already suggested by a knee in the light curve and a possible detection of broad H$\alpha$ emission at the same phase. We also analysed the sample of all 16 SLSNe-I with polarimetric measurements to date. The data taken during the early spectroscopic phase show consistently low polarisation indicating at least nearly spherical photospheres.  No clear relation between the polarimetry and spectral phase was seen when the spectra resemble Type Ic SNe during the photospheric and nebular phases. The light curve decline rate, which spans a factor of eight, also shows no clear relation with the polarisation properties. While only slow-evolving SLSNe-I have shown non-zero polarisation, the fast-evolving ones have not been observed at sufficiently late times to conclude that none of them exhibit changing $P$. However, the four SLSNe-I with increasing polarisation degree also have irregular light curve declines. For up to half of them, the photometric, spectroscopic and polarimetric properties are affected by CSM interaction. As such CSM interaction clearly plays an important role in understanding the polarimetric evolution of SLSNe-I.}

   \keywords{(Stars:) supernovae: general -- Polarization --  Techniques: polarimetric}

   \maketitle
%

\section{Introduction}


Superluminous supernovae (SLSNe) are an enigmatic class of stellar explosions that are characterised by exceptionally bright, often long-lived light curves \citep[e.g.][]{Gal-Yam2009, Pastorello2010, Chomiuk2011, Quimby2011}. While it has been nearly two decades since the first discovered events \citep[e.g. SN\,2005ap;][]{Quimby2007}, the mechanism to power their extreme luminosities is still a topic of debate \citep[for reviews, see e.g.][]{Gal-Yam2012, Moriya2018, Gal-Yam2019, Nicholl2021}. It is difficult to explain the photometric evolution of most SLSNe with the decay of radioactive \element[][56]{Ni} -- the canonical power source of Type Ia and Ibc supernovae (SNe) --  and alternative scenarios have been sought. The most popular models are either related to a spin-down of a highly-magnetised neutron star (magnetar) formed in the aftermath of the core-collapse  \citep[e.g. ][]{Woosley2010, Kasen2010} or to some manner of interaction between the SN ejecta and circumstellar material \citep[CSM; e.g.][]{Chevalier2011}, but the importance of jets has also been discussed \citep[e.g.][]{Soker2017, Soker2022}. However, despite the abundant data sets of both photometry \citep[e.g.][]{DeCia2018,Angus2019, Chen2022} and spectroscopy \citep[e.g.][]{Quimby2018}, there is no clear consensus on the most viable scenario. 

One way to gain further insight on the nature of the energy source of SLSNe is to investigate the evolution of the photospheric shape of the unresolved SNe via polarimetry. The continuum polarisation is induced by Thomson scattering of light from free electrons that are abundant in the SN ejecta \citep{Shapiro1982} and thus polarimetry directly probes the geometry of the photosphere. In case the SN has a circular projection on the sky, there is no prevalent direction for the polarised light and zero polarisation is expected. However, if the projection deviates from a perfect circle due to intrinsic asphericity of the photosphere, a non-zero polarisation signal is produced. For more detailed description of polarimetry in the context of SNe see \citet{Wang2008} and \citet{Patat2017}. For SLSNe, polarimetry is potentially a vital tool. In case they are powered by an internal engine such as a magnetar, the energy input is expected to be aspherical. As the ejecta expand over time, the electron density drops, and a more aspherical configuration of the inner layers of the ejecta should be revealed to the observer. The field of studying the geometry of SLSNe is still in its infancy. Only ten hydrogen-deficient SLSNe-I have any polarimetric data in the literature out of which only three have spectropolarimetry that provided concrete results.

Perhaps the best-studied case of SLSNe-I with polarimetry is SN\,2015bn. The slowly-evolving event was observed in both spectropolarimetry \citep{Inserra2016} and broad-band imaging polarimetry \citep{Leloudas2017a}. Two epochs of spectropolarimetry were obtained at $-24$ days before and $+27$ days after peak and follows a well-defined axial symmetry. The broad-band polarimetry spanned nine epochs between $-20$ and $+46$\,d with an increase of polarisation from $\sim 0.5\%$ to $>1\%$. For a hypothetical oblate spheroid photosphere which follows an inverse-square power law radial density distribution, the broad-band polarisation indicates an axial ratio of $>1.2$ \citep{Hoflich1991}. Both spectropolarimetry and broad-band polarimetry suggest an increased deviation from spherical symmetry toward the deeper layers of the ejecta. \citet{Leloudas2017a} proposed that the ejecta underwent a structural change at around $+20$\,d, when the photospheric emission shifted from an outer layer, dominated by C and O, to a more aspherical inner core, dominated by heavier freshly-synthesised material. This is supported by the simultaneous spectral change noted by \cite{Nicholl2016}. 

The slowly declining SN\,2017egm was observed near maximum light in spectropolarimetry \citep{Bose2018} and broad-band polarimetry \citep{Maund2019}. The spectropolarimetry, as re-analysed by \citet{Saito2020}, shows a polarisation of $\sim0.2\%$ near peak that was not strongly dependent on wavelength. They suggested a modest departure from spherical symmetry with an axial ratio of $\sim1.05$. \citet{Maund2019} did not detect intrinsic polarisation at four epochs spanning from $+4$\,d to $+19$\,d. \citet{Saito2020} report Subaru spectropolarimetry of SN\,2017egm at $+185$\,d, the only such measurement of a SLSN at this late stage. They find that the polarisation increased to $\sim0.8\%$, corresponding to an axial ratio $\sim1.2$. The late polarisation shows a nearly constant position angle over the wavelength range, suggesting an axisymmetric structure similar to that of SN\,2015bn, and \citet{Saito2020} conclude that the inner ejecta are more aspherical than the outer ejecta. 


Recently, \citet{Pursiainen2022} analyse two epochs of spectropolarimetry of SN\,2018bsz -- the most nearby SLSN-I to date at $z=0.0267$ \citep[see e.g.][]{Anderson2018, Chen2021} --  along an extensive spectroscopic data set. They conclude that the polarimetric and spectroscopic properties of SN\,2018bsz show clear evidence of the SN ejecta interacting with closed-by, highly aspherical CSM. After the explosion, the expanding ejecta quickly overtook the CSM and the first epoch of polarimetry at $+10$\,d was probing the geometry of the ejecta. However, as the photosphere receded, the aspherical CSM re-emerged as indicated by a multi-component H$\alpha$ line that appeared at $\sim+30$\,d. Therefore, the second epoch of polarimetry at $+40$\,d probed the photosphere that was strongly influenced by CSM resulting in a drastic change of the polarisation properties. While the authors could not determine the interstellar polarisation (ISP) induced by dichroic absorption of non-spherical dust grains partially aligned by the interstellar magnetic field, assuming that the ejecta at $+10$\,d is spherically symmetric, a $1.8\%$ rise of polarisation at $+40$\,d can been identified. Furthermore, \citet{Maund2021} analysed complementary imaging polarimetry and obtained one detection of $P\sim2.0\pm0.5\%$ at $+19\,$d.

There are also a number of SLSNe-I that have been observed exclusively with broad-band polarimetry. \citet{Leloudas2015b} presented imaging polarimetry of the fast-evolving SLSN-I LSQ14mo spanning from $-7$ to $+18$ days. The level of polarisation was constant at $\sim0.5\%$ and most likely attributable to the ISP. While this might be an indication of a small deviation from spherical symmetry, \citet{Inserra2016} noted that LSQ14mo was fainter than SN\,2015bn and it is possible that there was insufficient signal to noise (S/N) ratio in the data, and an asymmetric geometry like that of SN\,2015bn cannot be completely ruled out. Low level of broad-band polarisation have also been obtained for other SLSNe-I near peak brightness. \citet{Cikota2018} presented broad-band polarimetry of the fast-evolving PS17bek at four epochs between $-4$ and $+21$ days and no polarisation signal intrinsic to the SN was detected. \citet{Lee2020} reported an insignificant detection of polarisation in a single epoch of near-peak imaging polarimetry for SN\,2020ank. \citet{Lee2019} also shows that the $1.9\%$ polarisation of SN\,2018hti measured around its peak luminosity is consistent with the level of the ISP. On the other hand, \citet{Poidevin2022} presented imaging polarimetry of SN\,2020znr at $\sim+30$\,d and $\sim+290$\,d. Based on the non-detections at both epochs they suggest that the high ejecta mass of the exceptionally slowly evolving SLSN-I may have prevented the probe of to the inner geometry of the SN. Finally, \citet{Poidevin2022a} provide imaging polarimetry for two SLSNe-I. The single epoch of SN\,2021bnw at $+81$\,d showed no departure from symmetry, but the four epochs of SN\,2021fpl between $+2$\,d and $+43$\,d showed consistent polarisation around $1$\%. Based on spectral comparison, the authors concluded that SN\,2021fpl underwent a spectral transition similarly to SN\,2015bn but at an earlier phase, possibly explaining the consistently high polarisation.

Additionally, \citet{Cikota2018} presented the first circular polarimetry for two SLSNe-I. The fast-evolving PS17bek was observed around peak and the slow-evolving OGLE16dmu at $+100$\,d. Neither event showed evidence of circular polarisation. The non-detection did not constrain the magnetar scenario since a signal would only be expected close to the surface of the magnetar.
 
The scarce sample of polarimetric data for SLSNe-I shows that we do not have anywhere near a complete view of the time-evolving 3D structure of these spectacular events. While some events show distinct evidence for axisymmetry, others do not. It is not clear whether SLSNe-I with fast-evolving light curves have different geometry than the slow-evolving ones, nor whether the spectral phases of the polarimetry show any polarimetric tendency. Furthermore, the evidence that some SLSNe-I show greater asymmetry in deeper ejecta, as do standard core-collapse supernovae, has not yet been converted to quantitative evidence for an internal engine, such as a magnetar. Instead, CSM has been identified in a number of SLSNe-I \citep[e.g.][]{Yan2015, Yan2017, Lunnan2018, Pursiainen2022} and the irregular light curves seen in many SLSNe-I \citep[see e.g. ][]{Hosseinzadeh2021, Chen2022a} suggests that CSM interaction may play a role even in SLSNe-I that display no obvious spectral signatures attributed to CSM.  

Here we present polarimetric data for seven SLSNe-I of which only SN\,2017egm has previously published polarimetric observations. In particular, we present two epochs of spectropolarimetry for SN\,2017gci with complementary imaging polarimetry. We also perform a polarimetric sample analysis of SLSNe-I. In Section \ref{sec:obs} we present the observations and the data reduction procedures and in Section \ref{sec:isps} we discuss the ISP estimation. In Section \ref{sec:gci} we focus on the analysis of spectral and imaging polarimetry of SN\,2017gci and in Section \ref{sec:SLSN_pola} we investigate the sample properties of all SLSNe-I with polarimetry. We conclude our study in Section \ref{sec:conclusions}.



\section{Observations and Data Reduction}
\label{sec:obs}

Table \ref{tab:SLSN_list} provides an overview of the SLSNe-I with linear polarimetric observation to date. Only ten events are available in the literature and the sample we present in this paper increases the number of SLSNe-I with polarimetry by $\sim60\%$. For SN\,2017gci we provide both spectral and imaging polarimetry taken with the European Southern Observatory (ESO) Very Large Telescope (VLT; Section \ref{subsec:VLT_pola}). For SN\,2018ibb we present four epochs of VLT imaging polarimetry. The details of the reduction and in-depth analysis of the SN is presented in Schulze et al. (in prep). The five remaining SLSNe-I were observed in the imaging polarimetry mode at the Nordic Optical Telescope (NOT; Section \ref{subsec:NOT_pola}). All values of $P$ presented in this paper have been corrected for polarisation bias \citep[e.g.][]{Simmons1985, Wang1997} following \citet{Plaszczynski2014}.

\begin{table}
    \def\arraystretch{1.1}%
    \setlength\tabcolsep{10pt}
    \centering
    \fontsize{10}{12}\selectfont
    \caption{The 16 SLSNe-I with linear polarimetry. Peak MJD refers to the observed $r/V$ band peak. }
    \begin{threeparttable}
    \begin{tabular}{l c c c}
    \hline
    \hline
        \multicolumn{1}{c}{SN} & Peak MJD & $z$ & Refs.   \\
    \hline
    LSQ14mo                  & 56697.1    &  0.2560 & [2]                 \\
    SN\,2015bn               & 57102.5    &  0.1136 & [3,4]               \\
    PS17bek                  & 57814.6    &  0.3099 & [5]                 \\
    SN\,2017egm              & 57926.2    &  0.0310 & [\textbf{1},6,7,8]  \\
    SN\,2017gci              & 57986.8    &  0.0870 & [\textbf{1}]      \\
    SN\,2018bgv              & 58256.3    &  0.0790 & [\textbf{1}]     \\
    SN\,2018bsz              & 58267.5    &  0.0267 & [9,10]          \\
    SN\,2018ffj\tnote{\bf a} & 58337.6    &  0.2340 & [\textbf{1}]        \\
    SN\,2018hti              & 58462.1    &  0.0614 & [11]             \\
    SN\,2018ibb              & 58455.0    &  0.1660 & [\textbf{1},12]                \\
    SN\,2019neq              & 58734.0    &  0.1075 & [\textbf{1}]     \\
    SN\,2020ank              & 58898.5    &  0.2485 & [13]             \\
    SN\,2020tcw              & 59130.1    &  0.0645 & [\textbf{1}]        \\
    SN\,2020znr              & 59227.5    &  0.1000 & [14]                \\
    SN\,2021bnw              & 59264.5    &  0.0980 & [15]                \\
    SN\,2021fpl              & 59342.5    &  0.1150 & [15]                \\
    \hline
    \hline
  \end{tabular}
    \begin{tablenotes}
    \item[a] Caught on decline, MJD of first detection shown.

    \item[] \textbf{References:} 
    [1] This study; [2] \citet{Leloudas2015b}; [3] \citet{Inserra2016}; [4] \citet{Leloudas2017a}; [5] \citet{Cikota2018}; [6] \citet{Bose2018}; [7] \citet{Maund2019}; [8] \citet{Saito2020}; [9] \citet{Maund2021}; [10] \citet{Pursiainen2022}; [11] \citet{Lee2019}; [12] Schulze et al. (in prep); [13] \citet{Lee2020}; [14] \citet{Poidevin2022}; [15] \citet{Poidevin2022a}.
    \end{tablenotes}
    \end{threeparttable}    
\label{tab:SLSN_list}
\end{table}

\subsection{VLT Spectral and Imaging Polarimetry of SN\,2017gci}
\label{subsec:VLT_pola}

We observed SN\,2017gci with the FOcal Reducer and low dispersion Spectrograph (FORS2; \citealt{Appenzeller1998}) mounted on the Cassegrain focus of the VLT at Cerro Paranal in Chile, in both spectral (PMOS) and imaging polarimetric (IPOL) modes. Spectropolarimetry was obtained at two epochs, on 2017-08-25/26 and 2017-09-22 corresponding to $+3.3$ and $+29.0$ rest frame days relative to the peak brightness, respectively.  All observations were conducted with the 300V grism. Spectropolarimetry data was obtained using a $1\arcsec$ slit. The absence of the GG435 order-sorting filter is adopted intentionally to extend the wavelength coverage in the blue. Complementary imaging polarimetry was obtained at four epochs on 2017-08-29, 2017-09-23, 2017-10-19 and 2017-12-17, corresponding to $+7.0$, $+29.9$, $+53.8$ and $+108.0$\,d in rest frame with the $V_\mathrm{HIGH}$ FORS2 standard filter ($\lambda_0=555$\,nm, $\mathrm{FWHM}= 123.2$\,nm). All observations were obtained at four half-wave retarder plate (HWP) angles (0$\degree$, 22.5$\degree$, 45$\degree$, 67.5$\degree$) per cycle. The spectropolarimetry was reduced using standard procedures with \texttt{IRAF} \citep[for details see ][]{Cikota2017}. Furthermore, we used wavelet decomposition to reduce the noise of the flux spectra of individual ordinary and extraordinary beams \citep[$o$ and $e$ beams hereafter;][]{Cikota2019}. The resulting spectra were compared against the original ones to ensure that no systematic errors were introduced. The imaging polarimetry was also reduced using standard routines \citep[for details, see e.g.][]{Leloudas2015b,Leloudas2017a}. The log of the VLT observations is given in Table~\ref{tab:gci_obs_log}.


\subsection{NOT Imaging Polarimetry}
\label{subsec:NOT_pola}

\begin{figure*}
    \centering
    \begin{subfigure}[b]{0.492\textwidth}
        \centering
        \includegraphics[width=\textwidth]{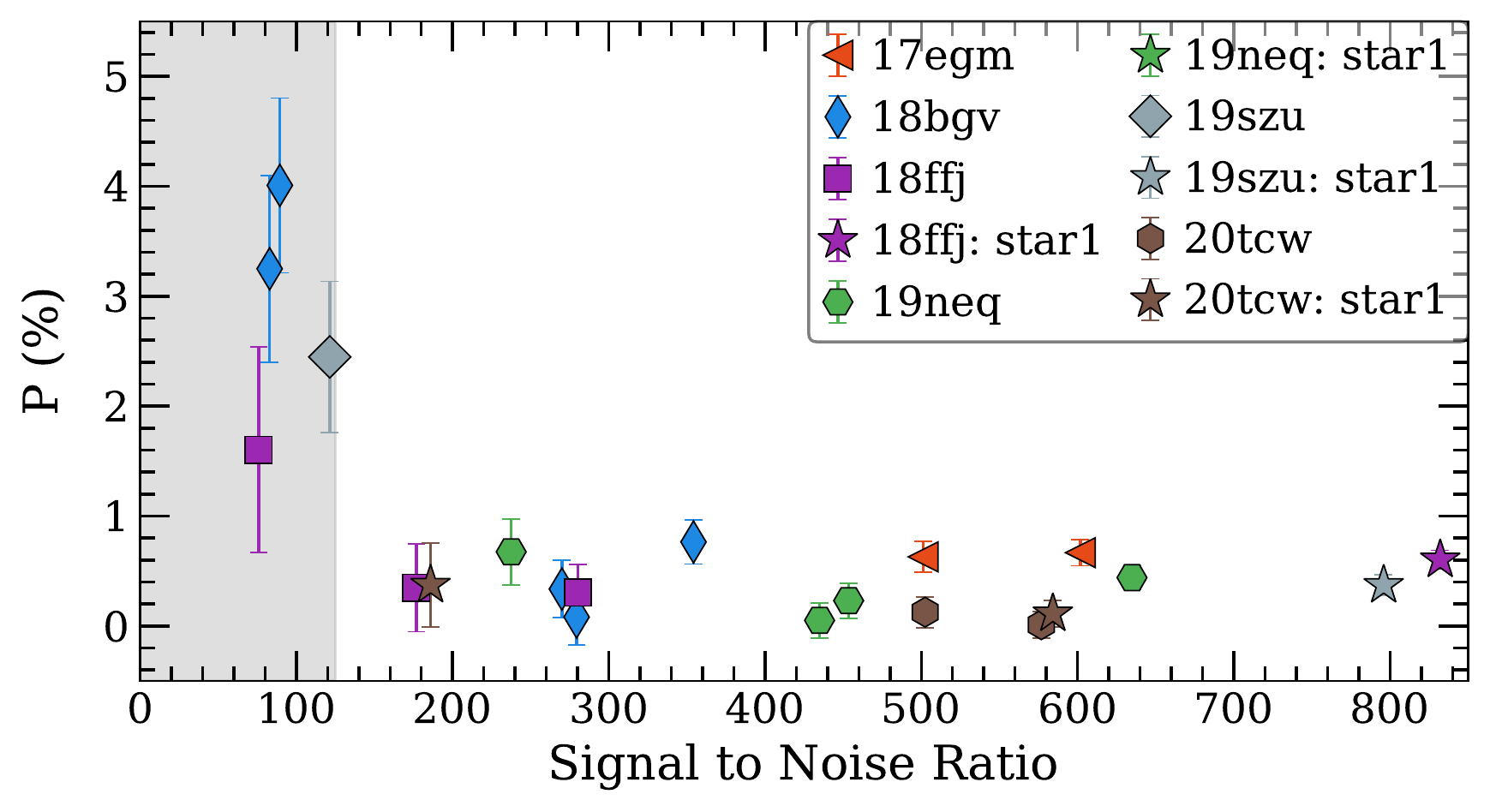}
    \end{subfigure} %
    \begin{subfigure}[b]{0.492\textwidth}
        \centering
        \includegraphics[width=\textwidth]{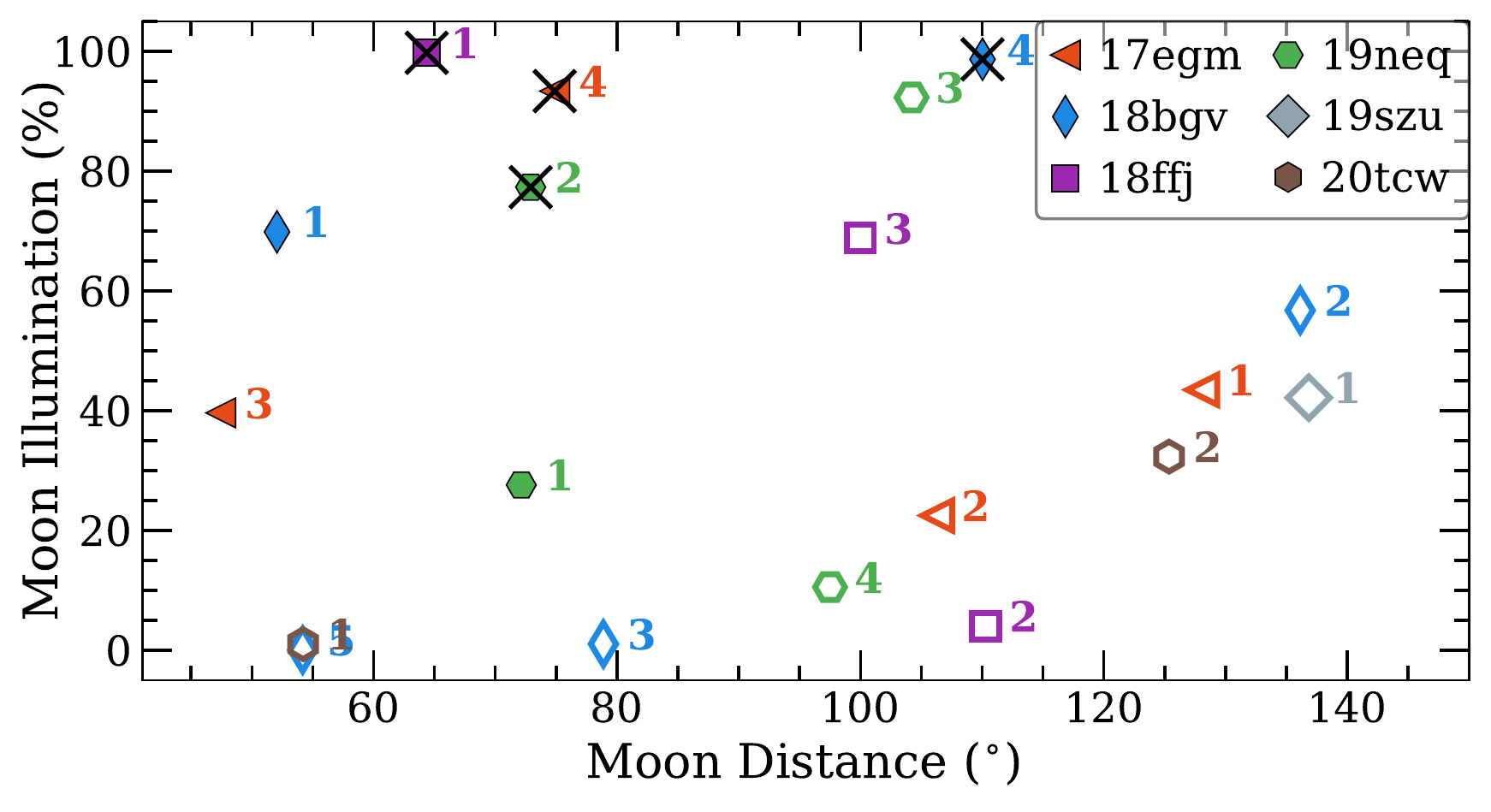}
    \end{subfigure} %

    \caption{\textit{Left}: The derived polarisation degree $P$ (\%) against the average S/N ratio of the target for the NOT/ALFOSC imaging polarimetry in the $V$ band. The $P$ values have been corrected for polarisation bias following \citet{Plaszczynski2014}. High values of polarisation degree are found only below S/N ratio $\sim125$ (grey region), implying these measurements are untrustworthy. Note that the single epoch of SN\,2019szu is also shown in the figure, but due to the low S/N ratio it is excluded from the analysis. \textit{Right}: The moon illumination against the SN -- moon separation at the time of the observations. The respective epoch of each SN is shown next to each marker. The epochs that were identified as questionable in comparison to neighbouring observations have been crossed in black.  Open markers denote epochs when the moon was below the horizon. The questionable observations were taken during high lunar illumination implying that it affects the results, rendering them unreliable.}
    \label{fig:P_vs_SNR}
\end{figure*}


We analyse imaging polarimetry of five SLSNe-I taken with the Alhambra Faint Object Spectrograph and Camera (ALFOSC) mounted on the 2.56\,m NOT at La Palma, Spain. Most observations were taken with $V$ band filter, but for SN\,2017egm, $R$ and $I$ were also used. Additionally, we observed SN\,2019szu approximately one month after its peak brightness. However, as we explain below the low signal to noise (S/N) ratio of the observation renders the results unreliable and the SN is excluded from the analysis.  All observations were obtained at four HWP angles (0$\degree$, 22.5$\degree$, 45$\degree$, 67.5$\degree$). The observation logs for the five SNe are presented in Tables \ref{tab:egm_obs_log} --  \ref{tab:tcw_obs_log}. 

For the NOT imaging polarimetry, a custom pipeline using \texttt{source extractor} \citep{Bertin1996} for the photometry was developed. The aperture sizes were determined based on the \texttt{source extractor} estimated FWHM. All images of an individual epoch (four images per cycle) were investigated and the smallest FWHM of the $o$ and $e$ beams was selected as the FWHM of that epoch. In each frame, the FWHM was estimated based on the SN and the bright stars in the field.

Considering the imaging polarimetry mode of ALFOSC tends to elongate the point-spread function of the sources in $o$ and $e$ beams differently \citep[see e.g.][]{Leloudas2017a}, the only feasible way to ensure that the same fraction of light was retrieved for every $e$ beam and $o$ beam source over the four angles, is to enclose majority of the target flux in the aperture. While smaller aperture size increase the S/N ratio and as such is ideal for photometry, the fraction of light in the $o$ and $e$ beams might be disproportionate thus inducing spurious polarisation in the observation. However, with larger aperture the amount of light from the target no longer increases but the S/N ratio decreases. In our experience aperture of $2$\,--\,$3\times\,\mathrm{FWHM}$ is optimal and in this paper all presented NOT imaging polarimetry values are obtain with aperture $2\,\times\,\mathrm{FWHM}$. We made extensive checks to verify the impact of the aperture size on the results. All results were found to be consistent with respect to the output of an alternative pipeline \citep{Leloudas2022} that utilises \texttt{DAOPHOT} \citep{Stetson1987}.

We find that the ALFOSC imaging polarimetry yields unreliable results under certain conditions. First, we find that the measured values of polarisation tend to be systematically higher when the S/N ratio of the target is $\lesssim125$ as shown in Figure \ref{fig:P_vs_SNR} (left), implying that these results are untrustworthy. This is likely a direct result of the low S/N ratio. As polarimetry is based on flux differences between the $o$ and $e$ beams, the higher uncertainty can induce unreal polarimetric signal. While the applied polarimetric bias correction should alleviate the issue, the corrections are statistical and designed assuming high S/N ratio and thus are not applicable. While it is possible that some of these observations truly show high polarisation, we exclude them from the analysis due to their uncertain nature. 

Furthermore, the data taken during high lunar illumination ($\sim75\%$) when the moon was above the horizon appear to be unreliable. We identified several observational epochs as questionable as the results show significant offset in comparison to neighbouring epochs for either the SN or a field star despite a high S/N ratio. As shown in Figure \ref{fig:P_vs_SNR} (right), these observations were taken during high lunar illumination, implying that it affects the observed polarisation signal. This is likely caused by Rayleigh scattering of the Lunar photons in the atmosphere, effect of which is the strongest at 90\degree. Any scattering process inherently induces linear polarisation perpendicular to the plane of scattering. As such the moon always creates polarised background emission and during bright moon it can dominate the polarised flux compared to the polarised signal extracted from the SN. Thus, unless the background is perfectly estimated in the analysis, the background polarisation can affect the measured values. The epochs highlighted in Figure \ref{fig:P_vs_SNR} (right) are: 

\begin{itemize}
    \item 1st epoch of SN\,2018ffj: The SN itself had a low S/N ratio ($\sim80$, see Table \ref{tab:impol_results} and Figure \ref{fig:QU_SNe}), but a comparison star in the field of view (FOV) has $\mathrm{S/N}\sim800$ and should be reliable. However, the Stokes $Q$ and $U$ parameters are clearly offset from the latter two measurements (see Figure \ref{fig:NOT_QU_stars}). 
    \item 2nd epoch of SN\,2019neq: The Stokes $Q$ and $U$ parameters are at an offset in comparison to the other epochs (Figure \ref{fig:QU_SNe}).
    \item 4th epoch of SN\,2017egm: Even the prominent host galaxy is not visible in the image rendering any analysis impossible.
    \item 4th epoch of SN\,2018bgv: High values of $Q$ and $U$ (Figure \ref{fig:QU_SNe}. Low S/N ratio likely (or at least partially) due to the high lunar illumination.
\end{itemize}

We also note that while the Stokes $Q$ and $U$ parameters at the first epoch of SN\,2018bgv are found to be similar to the two reliable epochs (see Figure \ref{fig:QU_SNe}), the observations were taken during high lunar illumination and are likely affected by it. Therefore we exclude this epoch from the analysis as well. Finally, we note that as low background level was measured when the bright moon was below the horizon, we consider the corresponding epochs reliable.

\section{ISP Correction}
\label{sec:isps}

\begin{figure*}
    \centering
    \includegraphics[width=0.98\textwidth]{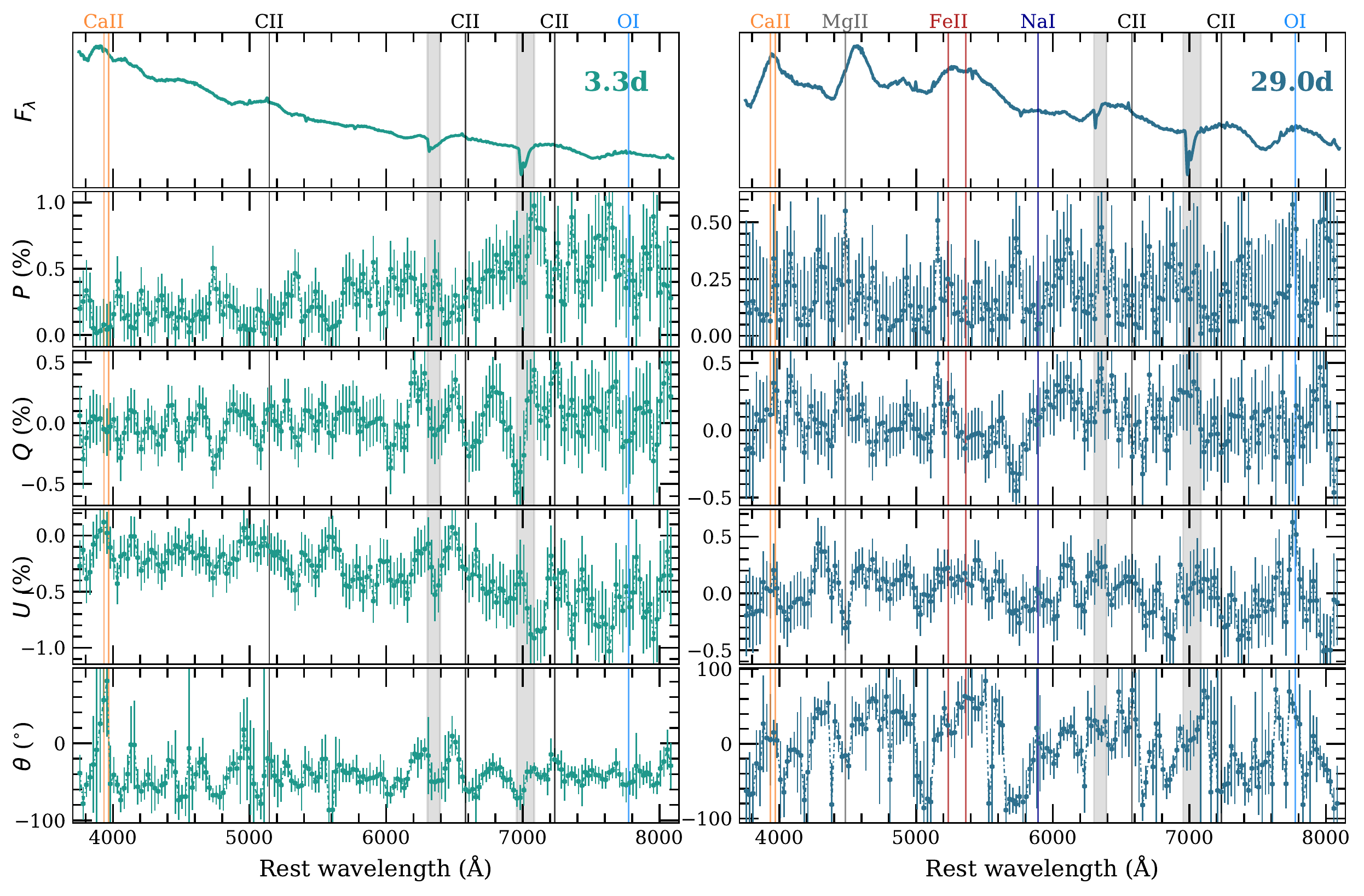}
    \caption{Flux spectrum and the ISP-corrected polarisation degree $P$, Stokes $Q$ and $U$ parameters and the polarisation angle $\theta$ for SN\,2017gci at $+3.3\,$d (left) and $+29.0$\,d (right). The polarimetry data are binned to 25\,Å in rest frame for visual clarity, but the flux spectra are shown at the natural binning of the spectrograph ($\sim3.3$\,Å). The most notable line features (vertical lines) and telluric bands (grey regions) are highlighted. At $+3.3$\,d the polarisation degree increases towards the red while at $+29.0$\,d it is found at a consistently low value.}
    \label{fig:FPQUX_specpol_gci}
\end{figure*}

\begin{table}
    \def\arraystretch{1.1}%
    \setlength\tabcolsep{6pt}
    \centering
    \fontsize{10}{12}\selectfont
    \caption{The ISPs used in this paper.}
    \begin{threeparttable}
    
    \begin{tabular}{l r r r}
    \hline
    \hline
    \multicolumn{1}{c}{SN} & \multicolumn{1}{c}{$Q_\mathrm{ISP}$ (\%)}  & \multicolumn{1}{c}{$U_\mathrm{ISP}$ (\%)}   & \multicolumn{1}{c}{$P_\mathrm{ISP}$ (\%)} \\
    \hline
SN\,2017gci               & $-0.50 \pm 0.01$ & $-0.02 \pm 0.01$ & $ 0.50 \pm 0.01$ \\
SN\,2017egm\tnote{\bf a}  & $ 0.29 \pm 0.34$ & $-0.40 \pm 0.38$ & $ 0.49 \pm 0.38$ \\
SN\,2018bgv\tnote{\bf b}  & $ 0.00 \pm 0.00$ & $ 0.00 \pm 0.00$ & $ 0.00 \pm 0.00$ \\
SN\,2018ffj               & $ 0.09 \pm 0.02$ & $-0.08 \pm 0.02$ & $ 0.12 \pm 0.02$ \\
SN\,2019neq               & $ 0.27 \pm 0.03$ & $-0.18 \pm 0.03$ & $ 0.32 \pm 0.03$ \\
SN\,2020tcw               & $-0.16 \pm 0.11$ & $-0.03 \pm 0.11$ & $ 0.17 \pm 0.11$ \\
    \hline
    \hline
    \end{tabular}
    \begin{tablenotes}
    \item[a] From \citet{Saito2020}.
    \item[b] Based on \citet{Heiles2000}.
    \end{tablenotes}
    \end{threeparttable}    
\label{tab:isps}
\end{table}

One of the most crucial steps in analysing polarimetric data is to perform a correction for \textit{interstellar polarisation}, introduced by dust grains between the target and the observer. Not only can it completely dominate the observed polarisation signature \citep[see e.g. ][]{Stevance2019}, but due to its vector nature it can both introduce and weaken the observed polarisation degree. Therefore, in order to investigate the intrinsic polarisation of SNe its effect has to be carefully estimated and removed. Unfortunately there is no unambiguous way to determine the ISP. A summary of the commonly used methods to estimate ISP has been provided by \citet{Stevance2020}. In this paper we estimate the ISPs by measuring them from the reliable, bright stars in our imaging polarimetry. As such we only estimate ISP of the Milky Way covered by the stars. We do not have the means to estimate the host galaxy ISP, but given we retrieve consistently low level polarisation for all of our SNe (see Table \ref{tab:impol_results}), the contribution from the respective host galaxies appears to be small. This is expected given that the host galaxies of SLSNe-I are typically found to be low-mass \citep[e.g.][]{Neill2011, Lunnan2014, Leloudas2015, Perley2016, Schulze2018}.  

The adopted ISPs are summarised in Table \ref{tab:isps}. For imaging polarimetry, the ISPs are vectorially subtracted from the obtained Stokes $Q$ and $U$ parameters. For the spectropolarimetry of SN\,2017gci, the correction is performed using an empirical Serkowski law: $p(\lambda)/p_{\mathrm{max}} = \exp{[-K \ln^2{(\lambda_{\mathrm{max}}/\lambda)}]}$, where $p_{\mathrm{max}}$ is the maximum polarisation at wavelength $\lambda_{\mathrm{max}}$ \citep{Serkowski1973}. We used the $V$ band ISP ($P_\mathrm{ISP}=0.50\%$) and commonly used $\lambda_{\mathrm{max}} = 5500$\,Å\ and $K = 1.15$ \citep{Serkowski1975} for the correction. For the details of the ISP estimation for individual SNe, see Appendix \ref{appsec:isp_details}.

\section{Polarimetry of SN\,2017gci}
\label{sec:gci}

\begin{figure*}
    \centering
    \includegraphics[width=0.98\textwidth]{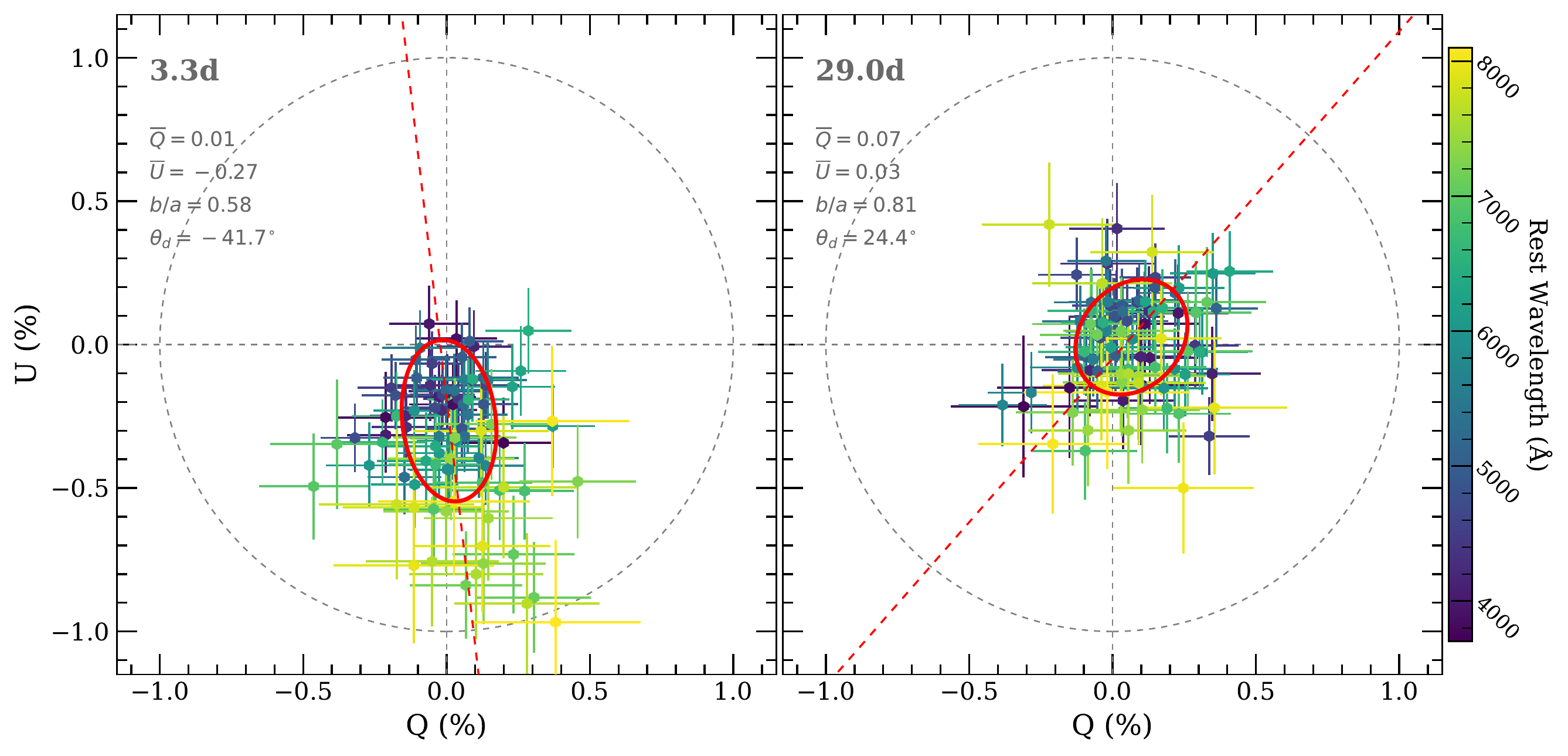}
    \caption{The $Q$\,--\,$U$ planes for the ISP-corrected spectropolarimetry of SN\,2017gci at $+3.3$\,d (left) and $+29.0$\,d (right). The data colours follow wavelengths as indicated by the colourbar and the dashed grey lines mark $Q = 0$, $U=0$ and $P=1$. The red ellipses drawn at the barycenter of the data ($\tilde{Q}$, $\tilde{U}$) are based on a principal component analysis \citep{Maund2010} and quantify the direction of the maximum variance of the polarisation (major axis) and the ratio of polarisation toward the orthogonal direction (the minor to major axis ratio $b/a$). Dominant axis (dashed red line) is drawn along the semi-major axis of the ellipse at an angle $\theta_d$ as measured from north towards east on the sky. At the first epoch the data appears to follow the dominant axis within the observational noise, but at the second epoch the points are clustered roughly circularly around $P=0$ with a weaker tendency along the drawn dominant axis. }
    \label{fig:QU_plane_specpol_gci}
\end{figure*}

In Figure \ref{fig:FPQUX_specpol_gci} we present flux spectra of SN\,2017gci along the ISP-corrected $P$, Stokes $Q$ and $U$ parameters and the polarisation angle $\theta$ as a function of wavelength for the VLT spectropolarimetry obtained at $+3.3$\,d and $+29.0$\,d. In both epochs we see a low level of polarisation along most of the continuum spectrum. However, there is one notable difference: at $+3.3$\,d the polarisation degree increases redward from $P\sim0.1$\,--\,$0.2\%$ at $\sim4000$\,Å to $P\sim0.5\%$ at $\sim8000$\,Å, while at $+29.0$\,d it is found at a constant level of $P\sim0.1\%$ throughout the wavelength range. 

The changes can also be seen in Figure \ref{fig:QU_plane_specpol_gci}, where we present the Stokes $Q$\,--\,$U$ planes for the ISP-corrected spectropolarimetry. Such diagrams offer an intuitive visualisation of the polarimetric properties as both $P$ ($\sqrt{Q^2 + U^2}$) and $\theta$ ($0.5\arctan(U/Q)$) are directly related to the Stokes parameters. Following \citet{Maund2010}, we use principal component analysis to define ellipses shown in red that encapsulate the spread of the data. The ellipses are drawn at the barycenter of the data ($\tilde{Q}$, $\tilde{U}$) and show the direction of the maximum variance at angle $\theta_d$ with respect to N/S axis on the sky ($\theta_d>0\degree$ implies tilt towards east). The shape of the ellipses also quantify the fraction of polarisation carried toward the dominant and orthogonal directions as the minor over major axial ratio ($b/a$). A ratio of $b/a=0$ refers to the ideal situation where all polarisation is carried by the dominant direction. To guide the eye we draw a \say{dominant axis} in the direction of the maximum variance $\theta_d$ in dashed red. At $+3.3$\,d the data are off-centre and follow the dominant axis at $\theta_d\sim-40\degree$ with $b/a=0.58$. At $+29.0$\,d, however, the data spread is nearly circular ($b/a=0.81$) and clustered around $P=0\%$. While we show a dominant axis at $\theta_d\sim25\degree$, the tendency of the data is significantly less prominent compared to $+3.3$\,d. We also note that as at the first epoch the dominant axis passes through $P=0\%$ and at the second epoch the data points are clustered around it, our estimate for the ISP appears to be correct.

A preferred orientation of data points on the Stokes $Q$\,--\,$U$ plane implies a presence of axial symmetry. Geometric deviation from spherical symmetry may be caused by three different mechanisms: i) inherently aspherical photosphere \citep[e.g.][]{Hoflich1991}, ii) absorbing body covering only a part of the photosphere \citep[e.g.][]{Kasen2003} and iii) an off-center power source \citep[e.g.][]{Hoflich1995}. Given that the low-level polarisation at $+29.0\,$d implies spherical symmetry, option iii) is unlikely the cause for the continuum polarisation at $+3.0$\,d. Furthermore, \citet{Fiore2021} identified broad \ion{C}{II} absorption lines in near-peak spectra of SN\,2017gci. As shown in Figure \ref{fig:QU_plane_loops_3d}, these lines do not project loops on the $Q$\,--\,$U$ plane -- as was seen in SN\,2018bsz \citep{Pursiainen2022} -- nor do they follow the drawn dominant axis as would be expected if the polarisation properties at $+3.3$\,d were caused by uneven distribution of absorbing material. Thus, option ii) does not seem to be the cause for the tendency of the data and we assume that the polarisation is caused by an inherently aspherical photosphere. The photosphere is found at angle of $\theta_d=-41.7\degree$ on the sky (i.e. westward from the N/S axis). For limiting polarisation of $\sim0.5\%$, the lower limit of the physical axial ratio is $a/b\sim1.1$ assuming an oblate ellipsoid \citep{Hoflich1991}. The data points appear to be mostly within the observational noise from the dominant axis, but we cannot rule out the possibility that some SN ejecta deviate from the discussed configuration inducing polarisation to the orthogonal direction. A signature of such deviations are the loops in the $Q$\,--\,$U$ plane caused by the absorbing material \cite[e.g.][]{Wang2008}. While we did not identify such loops, the data over \ion{C}{II} $\lambda6580$ appear to be tentatively orthogonal to the dominant axis but no such tendency is present in other lines (see Figure \ref{fig:QU_plane_loops_3d}). This could indicate a deviation from the slightly spheroidal configuration. However, as the data over the line does not pass through $P=0\%$, the position angle over it changes dramatically implying that the absorbing body would have to have a complicated structure in front of the photosphere. At the second epoch no clear loops are visible either. While only a weak tendency towards the dominant direction was seen in Figure \ref{fig:QU_plane_specpol_gci}, \ion{Mg}{II} $\lambda4481$ appears to be orthogonal to the dominant axis at $\theta_d\sim25\degree$ while \ion{Na}{I} $\lambda5893$ and tentatively \ion{C}{II} $\lambda6580$ are along it. While the \ion{Na}{I} $\lambda5893$ line is faint in the spectrum (see Figure \ref{fig:FPQUX_specpol_gci}), it becomes stronger soon after as shown by \citet{Fiore2021}.

\begin{figure*}
    \centering
    \includegraphics[width=0.98\textwidth]{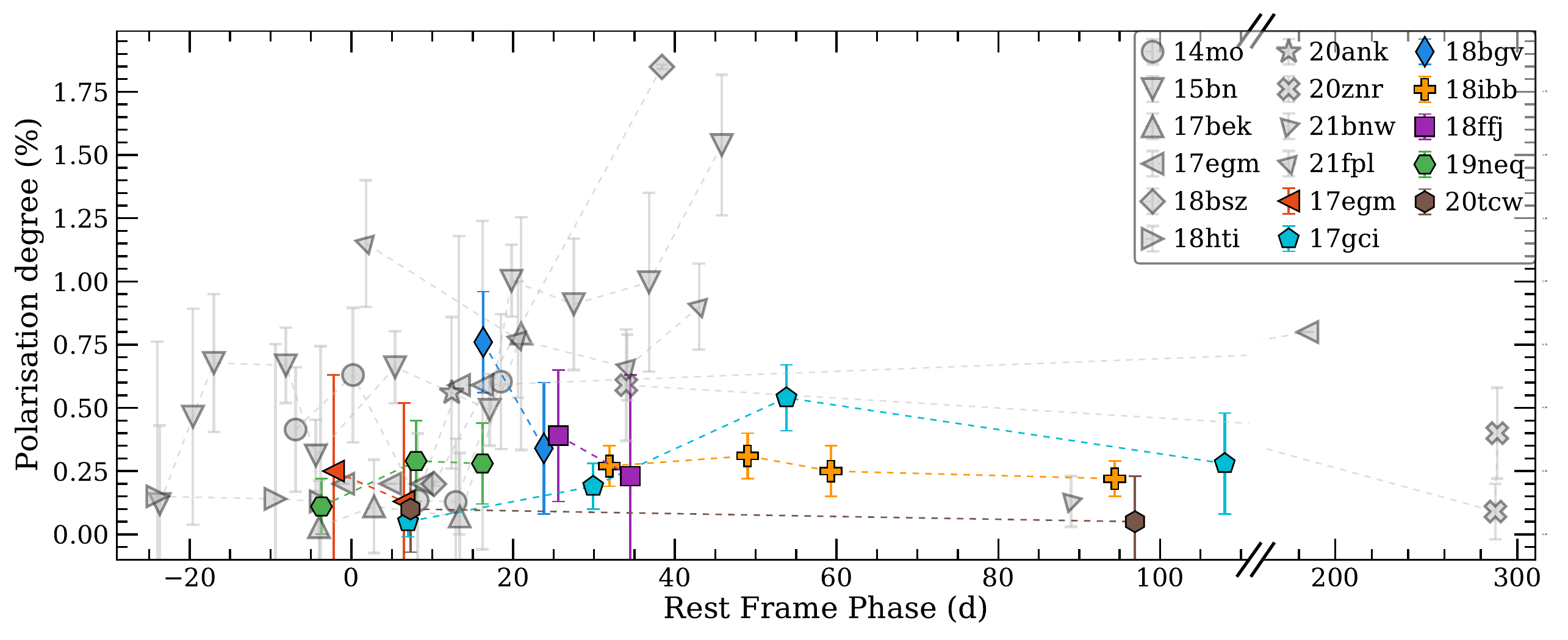}
    \caption{Polarisation degree vs. rest frame phase measured from the peak brightness in $r$/$V$ band. The literature data collected from sources listed in Table \ref{tab:SLSN_list} are shown with grey markers. The data are in $V$ band (or comparable) apart from SN\,2020znr ($R$). No ISP correction is performed for SN\,2018bgv. For SN\,2018hti we performed the ISP correction based on values presented in \citet{Lee2019} and for SN\,2018bsz we have assumed case A ISP as presented by \citet{Pursiainen2022}. Note the break and scale change in x-axis at 150\,d.}
    \label{fig:P_vs_t}
\end{figure*}

The two epochs of spectopolarimetry imply that the photosphere of SN\,2017gci evolved from a slightly aspherical configuration to a more spherical in the first month of its post-peak evolution. While the change appears to be modest it is still surprising given that SLSNe-I are typically observed to either have a roughly constant geometry or become more aspherical with time, and such evolution has not been seen in the other three SLSNe-I with spectropolarimetry to date (SN\,2015bn, SN\,2017egm and SN\,2018bsz). For SN\,2015bn the two epochs of VLT/FORS2 spectropolarimetry at $-23.7$\,d and $+27.5$\,d presented by \citet{Inserra2016}, show an increasing polarisation degree \citep[see also][]{Leloudas2017a}. The data also show a strong wavelength dependence at both epochs. In the first epoch, the polarisation degree appears to increase towards the red (as in SN\,2017gci) and at the second epochs towards the blue, following approximately the same dominant axis at both epochs. Based on Monte Carlo radiative transfer modelling, \citet{Inserra2016} propose that the wavelength dependency can be a result of increasing depolarisation towards the blue caused by an increase of line opacity from red-to-blue. A similar interpretation was made by \citet{Patat2012} to explain the blue-to-red increase of $P$ seen in spectropolarimetry of the subluminous Type Ia SN\,2005ke. For SN\,2017egm, \citet{Bose2018} presented three epochs of spectropolarimetry at $-1$\,d, $+5$\,d , $+9$\,d and \citet{Saito2020} a fourth epoch at $+185$\,d. The SN showed low level of polarisation at near-peak with no particular line features, but at $+185$\,d the level of continuum polarisation had increased to $\sim0.8\%$. Finally, for SN\,2018bsz the polarimetric evolution appears to be strongly affected by interaction with highly aspherical CSM \citep{Pursiainen2022}, and comparison to SN\,2017gci is not fruitful.

While the spectropolarimetry shows a low level of polarisation at $+29.0\,$d, the imaging polarimetry at $\sim+55$\,d shows that the polarisation degree has increased to $P\sim0.5\%$ in $V$ band. In the $Q$\,--\,$U$ plane the SN is found at $Q\sim-0.4\%$, $U\sim0.4\%$ (see Figure \ref{fig:QU_SNe}) -- in almost the opposite direction in comparison to $+3.3$\,d. Assuming that the $V$ band measurement represents the overall behaviour of the continuum polarisation, we find a position angle of $\theta \sim70^{\circ}$ at $+$55\,d, indicating a rotation of $\sim70^{\circ}$ on the sky compared to $+3.3$\,d. However, we note that the $V$ band covers several notable emission lines \citep[\ion{Fe}{II} in particular;][]{Fiore2021}, that are possibly depolarising the signal and the inherent continuum polarisation might be higher.

Interestingly, the maximum of $P$ coincides with two unusual observables discussed by \citet{Fiore2021}: a \say{knee} in the light curve and an appearance of a broad emission feature at $6520$\,Å that could be H$\alpha$ and related to interaction with CSM. Furthermore, \citet{Stevance2021} used the stellar models of Binary Population And Spectral Synthesis \citep[BPASS;][]{Eldridge2008, Eldridge2017, Stanway2018} code to investigate the progenitor properties, and conclude that the preferred progenitor -- a $30\,M_\odot$ star in a binary system -- can lose its hydrogen envelope via common envelope evolution, mass transfer in the binary system and strong stellar winds just prior to the explosion, explaining the H-poor nature of the SN but also the possible detection of H-rich CSM.  In case these three observables are related to each other, this is evidence that the SN showed CSM interaction that affected its photometric, spectroscopic and polarimetric evolution in a similar manner to SN\,2018bsz \citep{Pursiainen2022}, and might indicate a presence of aspherical CSM. Finally, the last data point at $\sim+108$\,d, shows a tentative decrease in the polarisation degree (see Figure \ref{fig:P_vs_t}). If real, this could signify the expected behaviour at late times when the SN is evolving towards the nebular phase. At this stage the optical depth of electron scattering -- and as a result the polarisation -- decreases as is already seen in, for example, Type IIP SNe \citep[e.g.][]{Leonard2006, Chornock2010}. However, due to the high uncertainty, it is unclear if the decrease is real.

We also compared the spectral and imaging polarimetry taken at comparable epochs. We retrieved the broad-band values from the spectral polarimetry by integrating the individual $o$ and $e$ beam spectra over the VLT $V_\mathrm{HIGH}$ filter. The resulting Stokes parameters are:
\begin{itemize}
    \item ~~$+3.3$\,d: $Q=0.01\pm0.02\%$, $U=-0.20\pm0.02\%$,
    \item $+29.0$\,d: $Q=0.02\pm0.02\%$,  $U=~~\,0.09\pm0.02\%$.
\end{itemize}
Based on the imaging polarimetry we find:
\begin{itemize}
    \item ~~$+7.0$\,d: $Q=-0.05\pm 0.06\%$,    $U=-0.05\pm0.06\%$,
    \item $+29.9$\,d: $Q=~~\,0.06\pm 0.09\%$,  $U=~~\,0.20\pm0.09\%$. 
\end{itemize}
The second epoch of spectral polarimetry ($+29.0$\,d) is consistent with imaging polarimetry taken the night after, but a minor difference is seen in the $U$ of the spectral polarimetry taken at $+3.3$\,d in comparison to $+7.0$\,d. This possibly signifies short timescale ($\sim4$\,d) variation, but could also be a consequence of different observational methods and reduction procedures. As such we consider the imaging and spectral polarimetry to yield consistent results.


\section{Evolution of Polarisation for a Sample of SLSNe-I}
\label{sec:SLSN_pola}

In Figure \ref{fig:P_vs_t} we present the evolution of polarisation degree as a function of the rest frame phase for all SLSNe-I with polarimetry to date. The seven SLSNe-I presented in this paper are highlighted with coloured markers, and the SNe found in the literature are shown in grey. For SN\,2017egm we show our data and literature data separately. As illustrated in the figure,  we increase the number of SLSNe-I with any polarimetric observations significantly, but we also present data at epochs $\gtrsim+100$\,d for two more SLSNe. 

\begin{figure*}
    \centering
    \includegraphics[width=0.98\textwidth]{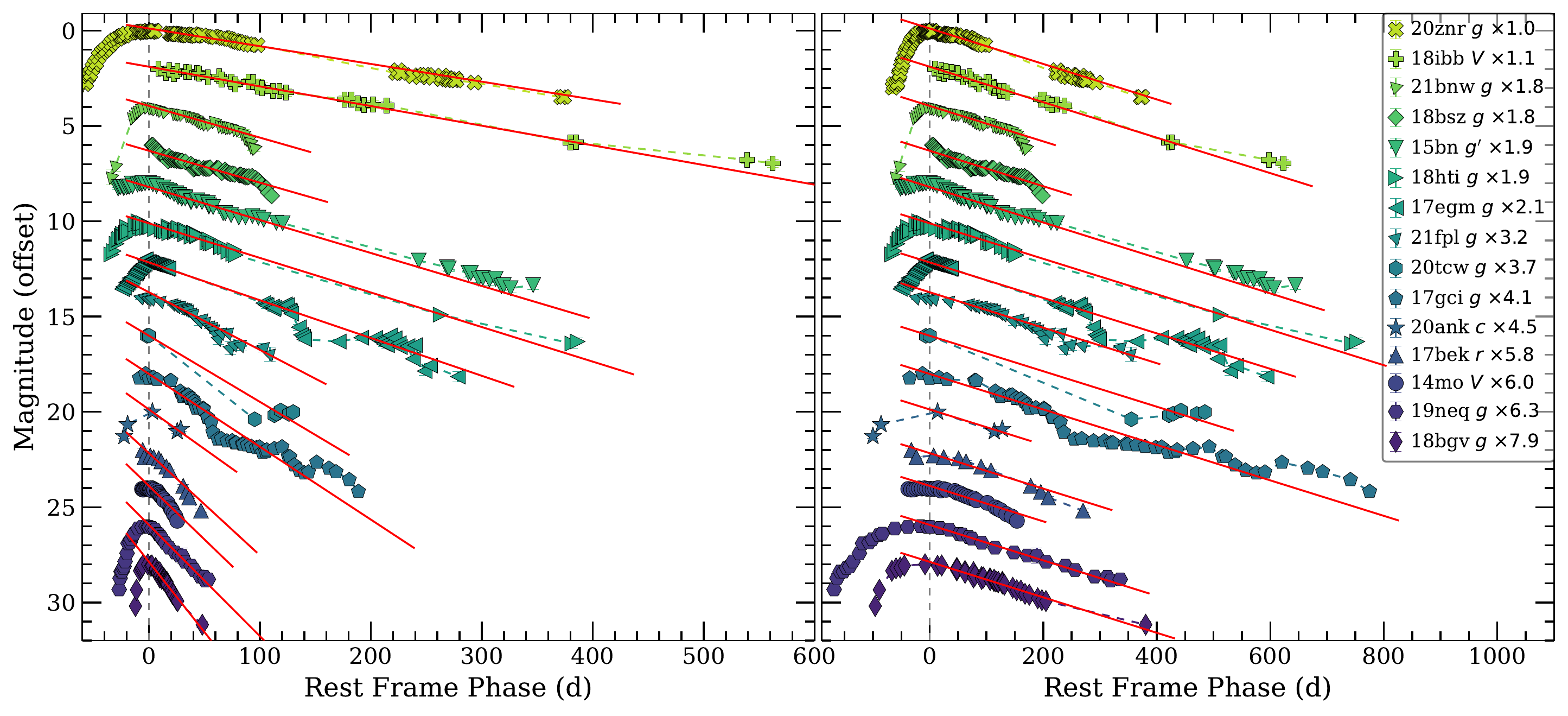}
    \caption{The light curves used to estimate the decline rates (left) and the light curves stretched to the decline rate of SN\,2020znr (right). The light curves are offset and the SNe are shown in the order from slow to fast for visual clarity. The used multiplication factors are given in the legend. The linear fits are shown with red lines. Only the data points after the dashed vertical grey line is used for the fits.  Note that SN\,2018ffj is excluded due to uncertain peak epoch. The light curves are collected from the following sources: SN\,2020znr, SN\,2021fpl, SN\,2020tcw and SN\,2019neq from ZTF via the Alerce broker \citep{Forster2021}, SN\,2018ibb (Schulze et al. in prep), SN\,2018bsz \citep{Chen2021}, SN\,2015bn \citep{Nicholl2016, Nicholl2016a}, SN\,2018hti \citep{Fiore2022}, SN\,2017egm \citep{Bose2018, Hosseinzadeh2021}, SN\,2017gci \citep{Fiore2021}, SN\,2020ank from ATLAS \citep{Tonry2018}, LSQ14mo \citep{Leloudas2015b}, SN2018bgv \citep{Lunnan2020}.}
    \label{fig:LC_slopes}
\end{figure*}

Considering the diverse light curve evolution timescales of SLSNe-I \citep[see e.g.][]{Nicholl2015a}, one-to-one comparison with respect to their peak times is not a reliable way to investigate the diversity of their polarimetric evolution. To account for this we have stretched the light curves so that they have a common decline rate using simple linear fits\footnote{The fits were performed using \texttt{LMFIT} \citep{Newville2014}.}. We chose SN\,2020znr as the reference as it is the slowest declining SLSNe-I in the sample. The light curve bands were chosen so that their central wavelengths would be similar, allowing a meaningful comparison. The details of the fits are shown in Table \ref{tab:decline_rates} and the light curves before and after the stretch in Figure \ref{fig:LC_slopes}. There is clearly a large diversity in the sample as the fastest declining SN in our sample (SN\,2018bgv) is 7.9 times faster than SN\,2020znr.

\begin{figure*}
    \centering
    \includegraphics[width=0.98\textwidth]{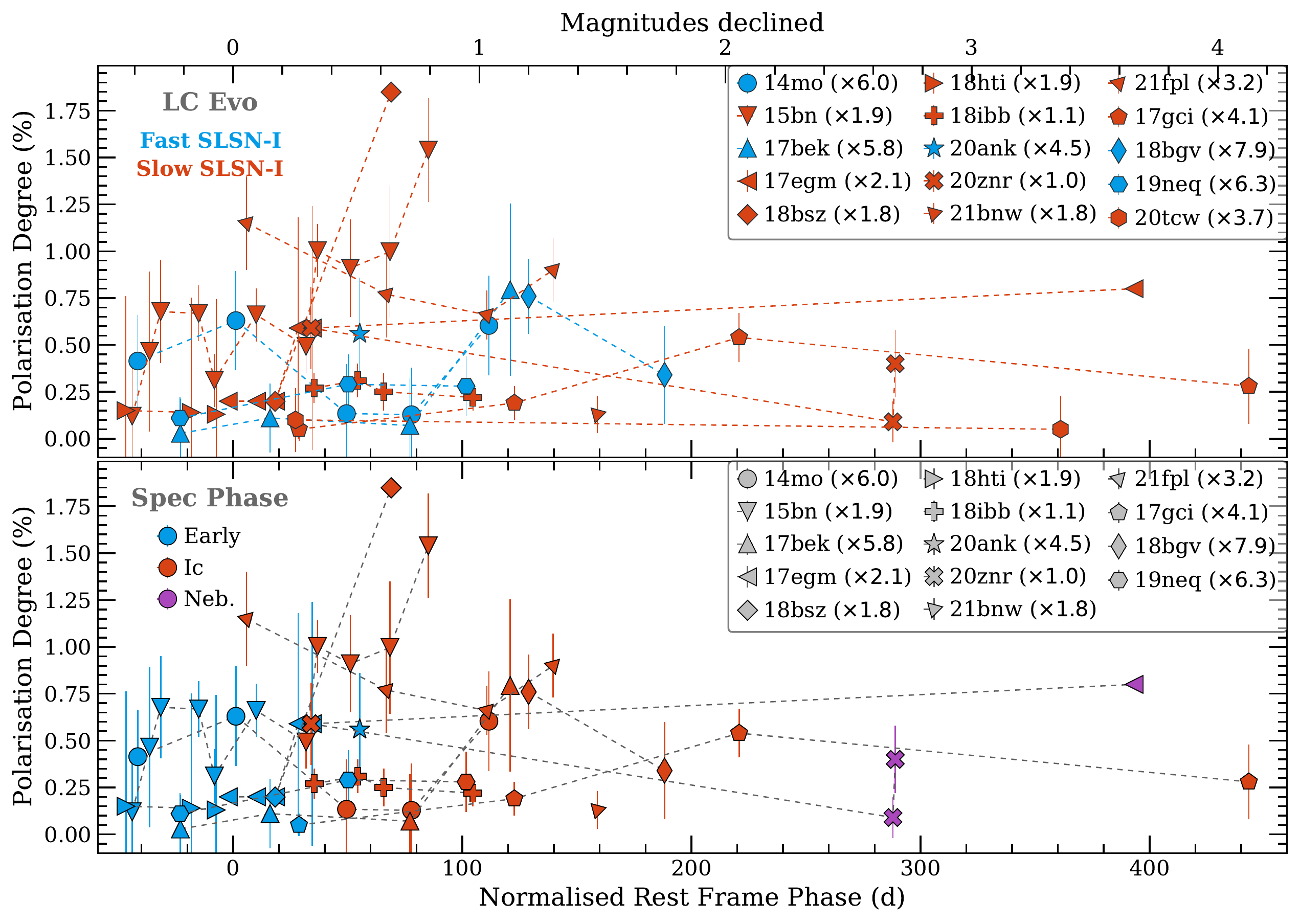}
    \caption{Polarisation degree vs. the normalised rest frame phase for SLSNe-I with polarimetric data highlighting the diverse decline rates (top) and spectroscopic phase of each epoch of polarimetry (bottom). The stretch factors used to normalise the decline rates are shown in the legend. Top axis refers to the number of magnitudes declined by the SNe, based on the linear fits. Only slow-evolving SLSNe-I have shown non-zero $P$, but the Fast ones have not been observed sufficiently late to rule it out. Spectroscopically, the SLSNe-I exhibit low polarisation during the Early phase, but no relation to polarimetry can be identified at the later phases. Note that SN\,2018ffj is not shown due to the unconstrained peak time and SN\,2020tcw is not shown in the bottom panel due to uncertain spectroscopic phases during the polarimetric epochs.}
    \label{fig:P_vs_t_spec_phase_lc_evo}
\end{figure*}

In the literature, SLSNe-I have been divided into \say{Fast} and \say{Slow} subgroups based on their photometric and spectroscopic properties \citep[see e.g.][]{Inserra2018a, Quimby2018}. In general, photometrically slow SLSNe-I also evolve spectroscopically slower in comparison to the fast-evolving ones. However, there is no clear threshold to separate the two based on their photometric decline rates. For the purposes of this analysis, we refer to the SNe that decline $\geq4$\,mag/100\,d as Fast SLSNe-I and the rest as Slow. While the choice is arbitrary, it agrees well with prototypical events such as LSQ14mo \citep[e.g.][]{Chen2017, Leloudas2015b} and SN\,2015bn \citep[e.g.][]{Nicholl2016}. Additionally, the classes of events such as SN\,2020ank \citep[Fast;][]{Kumar2021} and SN\,2017gci \citep[Slow;][]{Fiore2021} agree well with the Fast/Slow criteria of \citet{Inserra2018a}. In Figure \ref{fig:P_vs_t_spec_phase_lc_evo} (top), we present the polarisation degree against the normalised rest frame phase highlighting the Fast -- Slow distinction. Only Slow SLSNe have shown non-zero polarisation degree. However, there are only five Fast SLSNe and none of them were followed sufficiently late to conclude that they did not exhibit non-zero $P$. In fact, only SN\,2018bgv has been observed after it declined $\gtrsim1$\,mag from the peak brightness. Given that the slow-evolving SN\,2017gci showed increasing $P$ only after this it is not possibly to conclude if the Fast SLSNe do no the inherently exhibit non-zero polarisation degree.

\begin{table}
    \def\arraystretch{1.0}%
    \setlength\tabcolsep{5pt}
    \centering
    \fontsize{10}{12}\selectfont
    \caption{Details of the decline rate analysis. The stretch refers to the decline rate against SN\,2020znr. The SNe that decline faster than $4$\,mag/100\,d are called Fast.}
    \begin{threeparttable}
    \begin{tabular}{l c c c c c}
    \hline
    \hline
    \multicolumn{1}{c}{SN} &  \multicolumn{1}{c}{Band} & \multicolumn{1}{c}{$\lambda_\mathrm{eff}$ }  & \multicolumn{1}{c}{Rate}  & \multicolumn{1}{c}{Stretch} & \multicolumn{1}{c}{Class}     \\
    \multicolumn{1}{c}{} &  \multicolumn{1}{c}{ } & \multicolumn{1}{c}{(Å)}  & \multicolumn{1}{c}{(mag/100\,d)} & \multicolumn{1}{c}{}     \\
    \hline
LSQ14mo        & $V $ &  $4389$ & $5.62$ & $6.03$ & Fast  \\
SN\,2015bn     & $g'$ &  $4236$ & $1.73$ & $1.86$ & Slow  \\
PS17bek        & $r $ &  $4914$ & $5.36$ & $5.76$ & Fast  \\
SN\,2017egm    & $g $ &  $4660$ & $1.98$ & $2.13$ & Slow  \\
SN\,2017gci    & $g $ &  $4420$ & $3.82$ & $4.11$ & Slow  \\
SN2018bgv      & $g $ &  $4453$ & $7.36$ & $7.91$ & Fast  \\
SN\,2018bsz    & $g $ &  $4680$ & $1.67$ & $1.80$ & Slow  \\
SN\,2018hti    & $g $ &  $4527$ & $1.81$ & $1.95$ & Slow  \\
SN\,2018ibb    & $V $ &  $4727$ & $1.03$ & $1.11$ & Slow  \\
SN2019neq      & $g $ &  $4345$ & $5.84$ & $6.28$ & Fast  \\
SN2020ank      & $c $ &  $4269$ & $4.15$ & $4.46$ & Fast  \\
SN\,2020tcw    & $g $ &  $4514$ & $3.47$ & $3.73$ & Slow  \\
SN\,2020znr    & $g $ &  $4368$ & $0.93$ & $1.00$ & Slow  \\
SN\,2021bnw    & $g $ &  $4376$ & $1.66$ & $1.79$ & Slow  \\
SN\,2021fpl    & $g $ &  $4309$ & $3.02$ & $3.25$ & Slow  \\
    \hline
    \hline
    \end{tabular}
    \end{threeparttable}    
\label{tab:decline_rates}
\end{table}

Furthermore, \citet{Quimby2018} introduced the concept of spectroscopic phase for SLSNe-I. While it is beyond the means of this paper to replicate their procedure, we inspected during what kind of a spectral phase the polarimetric observations were obtained and divided them roughly into three categories: 1) the \say{Early phase} spectra consists of hot, blue continua that show mostly \ion{O}{II} absorption lines and/or \ion{C}{II} P\,Cygni profiles, 2) the \say{Ic phase} spectra are dominated by the elements seen in Type Ic SNe during the photospheric phase (e.g. Mg, Ca, O, Fe) and 3) \say{Pseudo-nebular} spectra are in general similar to nebular phase spectra of Type Ic SNe and show prominent forbidden emission lines (e.g. [\ion{Ca}{II}]). The spectroscopic phases are highlighted in Figure \ref{fig:P_vs_t_spec_phase_lc_evo} (bottom). No SLSN-I shows significant polarisation during the Early phase, directly implying that the outermost layers of SLSNe-I are consistently, at least nearly, spherical. While the spectral transition between Early and Ic phases results in an increase of polarisation for some events (e.g. SN\,2015bn), it does not appear to be the case for all (e.g. LSQ14mo). Finally, pseudo-nebular phase polarimetry has been obtained only for two SLSNe-I:  SN\,2020znr does not show increasing polarisation, but SN\,2017egm does. Thus, we conclude that the SLSNe-I exhibit low polarisation during the Early phase, but no relation between polarimetric properties and the spectroscopic phases were identified at the later phases.

Given that the sample of SLSNe-I with polarimetry at late times is still small it is not necessarily surprising that no strong tendencies were identified. One possibly important factor -- that we can not properly investigate due to the small sample especially at the late times -- is the effect of the viewing angle. The observed polarisation degree of a SN is effectively dictated by how circular the projection of the photosphere appears to be on the sky. While a small polarisation degree will always be obtained for an unobstructed spherical photosphere, an inherently aspherical one might result in a high or low value depending on the angle. As such the diversity of $P$ shown in Figure \ref{fig:P_vs_t_spec_phase_lc_evo}, could at least partially be cause by the distribution of the viewing angles. However, it is interesting to note that comparable or even smaller samples of other types of SNe have lead to strong conclusions on their nature. For instance, \citet{Chornock2010} showed that a sample of five Type IIP SNe was sufficient to verify that while the SNe have nearly spherical hydrogen envelopes (i.e. low $P$), their inner cores are highly aspherical. Therefore, the diversity of the polarisation properties of SLSNe-I might imply that the distribution of their photospheric shapes with time is inherently diverse.

Not only are the SLSNe-I with non-zero polarisation degree all slow-evolving ones, the four SNe that show increasing $P$ at some stage have all been reported to exhibit some manner of undulations or breaks during the light curve declines, while the other SLSNe-I with multi-epoch polarimetry show smooth photometric evolution (see Figure \ref{fig:LC_slopes}). The four SNe are SN\,2015bn \citep{Nicholl2016}, SN\,2017egm \citep{Hosseinzadeh2021}, SN\,2017gci \citep{Fiore2021}, SN\,2018bsz \citep{Chen2021}. While there are a few borderline cases where the topic is difficult to investigate due to poorly-sampled polarimetric data (SN\,2020tcw and SN\,2021bnw) or unclear polarimetric evolution (SN\,2021fpl), this implies that the irregular light curve evolution and the increasing polarisation degree are related. 

The irregular light curves have been a topic of discussion in the literature. In particular, \citet{Hosseinzadeh2021} and \citet{Chen2022a} investigated the prevalence of light curve decline undulations in SLSNe-I. While both conclude that they are common, they could not determine whether they are caused by CSM interaction or by a magnetar engine via an increase in energy input or a sudden decrease in the ejecta opacity. For SN\,2018bsz, \citet{Pursiainen2022} showed that the polarisation properties changed drastically after the CSM-related multi-component H$\alpha$ emission line appeared at $\sim+30$\,d. The light curve also entered a pseudo-plateau phase roughly at the same time, lasting until $\sim+100$\,d when one of the H$\alpha$ components suddenly faded. The authors concluded that both the polarisation properties and the light curve evolution were strongly affected by interaction with close-by, highly aspherical CSM that was enshrouded by the SN ejecta after the explosion and re-emerged only when the photosphere receded enough. For SN\,2017gci, \citet{Fiore2021} discuss the possibility that a broad spectral feature at $6520$\,Å -- which appeared at the same time as a knee in the light curve ($\sim+50$\,d) -- is H$\alpha$ emission arising from CSM. Given that we report an increase in polarisation degree at a similar phase ($\sim55$\,d), these observables are likely related and caused by the CSM interaction. On the other hand, the dense spectroscopic coverage of SN\,2015bn \citep{Nicholl2016, Nicholl2016a} and SN\,2017egm \citep{Nicholl2017d,Bose2018} does not reveal any obvious lines attributed to CSM interaction. However, \citet{Wheeler2017} model the bolometric light curve of SN\,2017egm up to $\sim+30$ d (i.e. before the light curve undulations) and argue that CSM interaction models can account for the sharp peak seen in the light curve, whereas the magnetar models have difficulties explaining it. Unfortunately, the sparse polarimetric data of SN\,2017egm does not allow an investigation on whether the change in polarisation occurred at the same time as the light curve undulations, but for SN\,2015bn the maximum polarisation is measured during a knee in the light curve. In summary, increasing polarisation degree and irregular light curve evolution appear to be related and in up to two out of four SLSNe-I with increasing $P$, photometric, spectroscopic and polarimetric properties are affected by CSM interaction. Therefore, we conclude that CSM interaction plays a key role in understanding the polarimetric evolution of  SLSNe-I.

Finally, we emphasise that the number of SLSNe-I with crucial late-time polarimetry is still very small. As shown in Figure \ref{fig:P_vs_t_spec_phase_lc_evo}, only three SNe have been observed after they have declined more than $2$\,mag. Based on the figure the spectra start to evolve towards the nebular phase only after this. As the photospheres recede over time, multi-epoch polarimetry from early times to such late phases are needed to map out the degree of asphericity at different layers of the ejecta, therefore providing a more comprehensive picture of the geometry of SLSNe-I. Thus, to investigate the true diversity of their polarimetric evolution and to test the different models for the underlying energy source, well-cadenced observations have to be obtained for more SLSNe-I both with and without irregular light curve evolution.

\section{Conclusions}
\label{sec:conclusions}

We have presented linear polarimetry for seven hydrogen-poor superluminous supernovae (SLSNe-I) of which only one has previously published polarimetric data. This is the largest dataset presented to date and it increases the number of SLSNe-I with polarimetry in the literature by 60\%. 
The best-studied event is SN\,2017gci, for which we presented two epochs of spectropolarimetry, previously obtained for only three SLSNe-I, along with further imaging polarimetry with VLT/FORS2. Using our data set and all available data in the literature we have also presented a sample analysis of SLSN-I polarimetry.

The analysis of SN\,2017gci revealed that it is the first SLSN-I for which a decrease of polarisation degree has been observed. In the first epoch of spectropolarimetry at $+3.3$\,d, the data are off-center on the $Q$\,--\,$U$ plane and follow a dominant axis  -- indicative of axial symmetry. Assuming an oblate photosphere, $P\lesssim0.5\%$ indicates a aspherical configuration with a physical axial ratio of $a/b\lesssim1.1$ \citep{Hoflich1991} if viewed equator-on. The polarisation degree increases from blue to red, which in literature has been attributed to be an effect of increasing line opacity towards the blue \citep[e.g.][]{Patat2012, Inserra2016}. The spectropolarimetry at $+29.0$\,d shows that the data is clustered around $P=0\%$ on the $Q$\,--\,$U$ plane, implying that the photosphere of SN\,2017gci became more spherical in the first month post-peak. The later imaging polarimetry shows that the polarisation degree increased again reaching a maximum of $P\sim0.5\%$ at $\sim+55$\,d, but towards a different direction on the $Q$\,--\,$U$ plane indicating a rotation of $\sim70\degree$ on the sky in comparison to $+3.3$\,d. The rotation might be related to CSM interaction as already implied by a simultaneous \say{knee} in the light curve and the possible detection of broad H$\alpha$ emission \citep{Fiore2021}.

Furthermore, we examined whether the polarimetric properties of SLSNe-I showed any relations with their photometric and spectroscopic properties. After normalising the light curves for the diverse decline rates, the polarimetry showed no clear relation with the photometric evolution timescale. While only Slow SLSNe-I have shown non-zero polarisation degree, the fast-evolving ones have not been observed at sufficiently late times to conclude that none of them show increasing $P$. Furthermore, the spectroscopic phases of the polarimetric epochs appear to have only a small effect. While the polarimetry taken during early spectroscopic phases (i.e. dominated by blue continuum) show low polarisation indicative of high spherical symmetry, no clear correlation between the polarimetric and spectroscopic evolution was seen at later epochs when the SNe were spectroscopically similar to Type Ic SNe during the photospheric or nebular phases.

We report that the four SLSNe-I that have shown increasing polarisation degree to date (SN\,2015bn, SN\,2017egm, SN\,2017gci and SN\,2018bsz) also exhibit irregular light curve declines, while the other SLSNe-I with multi-epoch polarimetry show smooth photometric evolution. Given that CSM interaction affected the spectroscopic, photometric and polarimetric properties of up to half of the four SNe, we conclude that the CSM interaction clearly plays an important role in understanding the polarimetric evolution of SLSNe-I. However, due to the small number of SLSNe-I with well-cadenced polarimetry up to late times, the effect of a possible internal engine cannot be properly investigated. More SLSNe-I both with and without light curve undulations need to be followed with polarimetry from near-peak to late times to probe the intrinsic diversity of the geometric structure of their ejecta.

\begin{acknowledgements}

We thank the anonymous referee for the helpful feedback.

M.P and G.L. are supported by a research grant (19054) from VILLUM FONDEN. 

A.C. is supported by NOIRLab, which is managed by the Association of Universities for Research in Astronomy (AURA) under a cooperative agreement with the National Science Foundation.

M.B. acknowledges support from the European Union’s Horizon 2020 Programme under the AHEAD2020 project (grant agreement n. 871158).

S.S. acknowledges support from the G.R.E.A.T. research environment, funded by {\em Vetenskapsr\aa det},  the Swedish Research Council, project number 2016-06012.

The research of Y.Y. is supported through a Bengier-Winslow-Robertson Fellowship. 
MN and AA are supported by the European Research Council (ERC) under the European Union’s Horizon 2020 research and innovation programme (grant agreement No.~948381) and MN also by a Fellowship from the Alan Turing Institute.

AGY’s research is supported by the EU via ERC grant No. 725161, the ISF GW excellence center, an IMOS space infrastructure grant and a GIF grant, as well as the André Deloro Institute for Advanced Research in Space and Optics, The Helen Kimmel Center for Planetary Science, the Schwartz/Reisman Collaborative Science Program and the Norman E Alexander Family M Foundation ULTRASAT Data Center Fund, Minerva and Yeda-Sela;  AGY is the incumbent of the The Arlyn Imberman Professorial Chair.

This work is based (in part) on observations collected at the European Organisation for Astronomical Research in the Southern Hemisphere under ESO programmes 099.D-0169(A) and 0100.D-0209(A) and on observations made with the Nordic Optical Telescope, owned in collaboration by the University of Turku and Aarhus University, and operated jointly by Aarhus University, the University of Turku and the University of Oslo, representing Denmark, Finland and Norway, the University of Iceland and Stockholm University at the Observatorio del Roque de los Muchachos, La Palma, Spain, of the Instituto de Astrofisica de Canarias under NOT programmes 57-010, 58-005, 59-013, 60-027, 62-003, 63-006. The NOT data presented here were obtained with ALFOSC, which is provided by the Instituto de Astrofisica de Andalucia (IAA) under a joint agreement with the University of Copenhagen and NOT.

This work benefited from L.A.Cosmic \citep{VanDokkum2001}, IRAF \citep{Tody1986}, PyRAF and PyFITS. PyRAF and PyFITS are products of the Space Telescope Science Institute, which is operated by AURA for NASA. We thank the authors for making their tools and services publicly available.  

\end{acknowledgements}

\bibliographystyle{aa} 
\bibliography{bib} 


\begin{appendix}

\section{ISP estimation}
\label{appsec:isp_details}
\subsection{SN\,2017egm}

As shown in Figure \ref{fig:NOT_stamps}, there is no star in the FOV to reliably reduce the ISP from our data. Therefore, we adopt the ISP of \citet{Saito2020} ($Q_{\rm ISP}=0.29\pm0.34\%$ and $U_{\rm ISP}=-0.40\pm0.3$\%) estimated based on the intrinsically unpolarised \ion{Ca}{II} NIR triplet emission present in the pseudo-nebular phase spectropolarimetry.

\subsection{SN\,2017gci}
\begin{figure*}[h]
    \centering
    \includegraphics[width=0.98\textwidth]{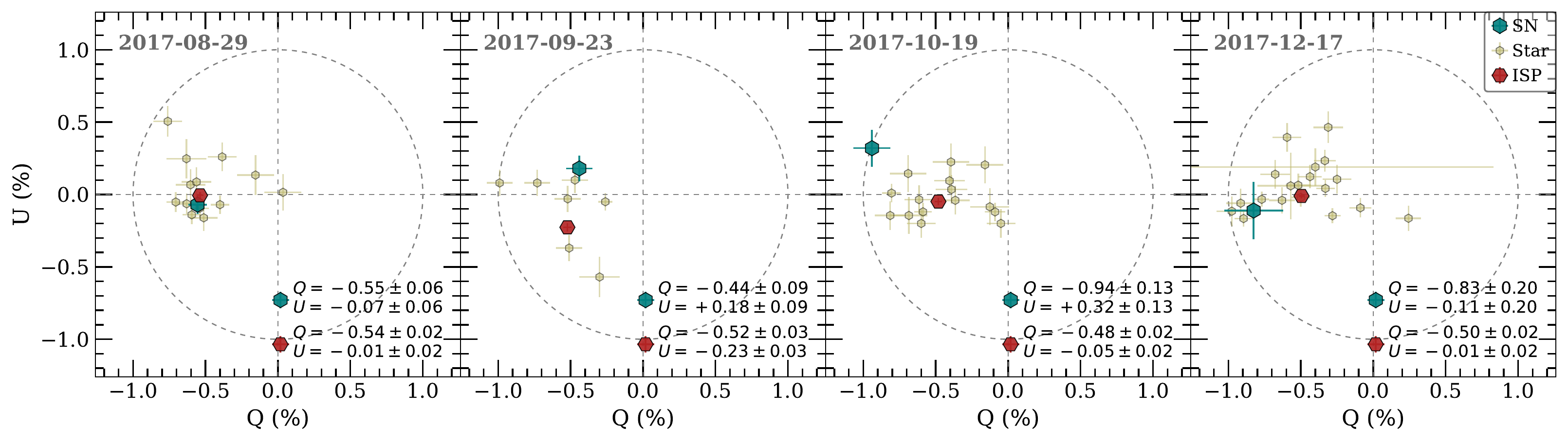}
    \caption{Stokes parameters of SN\,2017gci and bright field stars measured at four epochs displayed on the $Q$\,--\,$U$ plane. The SN (teal) and the field stars passing the selection criteria (olive) as well as derived Galactic ISPs (red) are shown. The second epoch consists only one cycle of observations, making the ISP estimate less reliable. For the other three epochs the derived values of ISP are consistent.}
    \label{fig:gci_QU_stars}
\end{figure*}

We estimated the ISP for SN\,2017gci using the numerous point sources in the VLT/FORS2 broad-band polarimetry. We used the three epochs with more than one cycle (set of observations over the four HWP angles), in order to exclude outliers.  We set a strict S/N ratio cut of 300 in every image to ensure accurate ISP caused by Galactic dust. First we estimated the ISP independently for each epoch by excluding measurements of stars that were not consistent between the cycles. For epochs one and three with two cycles, the field stars that have Stokes parameters that differ by $>1\sigma$ were excluded, while for epoch four with three cycles the remaining consistent measurements (if they existed) were kept. The ISP level for each epoch has been determined by using the error-weighted mean Stokes parameters and the final ISP by calculating the error-weighted average values among the three epochs of observations. Table \ref{tab:isp_gci} summarises the ISPs estimated for SN\,2017gci and we show the stars passing our criteria on the $Q$\,--\,$U$ plane in Figure \ref{fig:gci_QU_stars}. Note that we also show the results for epoch two, with only one cycle of observations, for which the ISP was estimated based on all the field stars that passed the S/N ratio cut. The large deviation in $U$ is likely to be caused by variations of the observing condition during the exposure, which cannot be cross-checked using multiple cycles of observations since only one set of polarimetry was obtained. We have also verified that the used stars are sufficiently far away to fully probe the Milky Way dust content. Following \citet{Tran1995}, a star needs to be $\sim150$\,pc above the Galactic plane to provide a good estimate for the Galactic ISP. Using the astrometric solutions from Gaia Data Release 3 \citep{GaiaCollaboration2022}, we conclude that all of the stars are found above the threshold and thus should yield reliable results.

\subsection{SN\,2018bgv}
We identify no bright point source in the FOV of the NOT/ALFOSC imaging polarimetry of SN\,2018bgv. (see Figure \ref{fig:NOT_stamps}). Instead, we refer to the polarisation standard stars published in \citet{Heiles2000} that are close to the location of SN\,2018bgv. There are two stars within 2$\degree$ from the SN and a further two within 5$\degree$, and the polarisation levels published for all these four stars are consistent with zero. Therefore, we adopt $Q_\mathrm{ISP}=0.0\%$ and $U_\mathrm{ISP}=0.0\%$.

\subsection{SN\,2018ffj}
For SN\,2018ffj, there is one bright star in the NOT/ALFOSC FOV (see Figure \ref{fig:NOT_stamps}) adequate to estimate the ISP. After excluding the first epoch due to high lunar illumination (see Section \ref{subsec:NOT_pola}), we find $Q_\mathrm{ISP}=0.09\pm0.02\%$ and $U_\mathrm{ISP}= -0.08\pm0.02\%$. The comparison with the Heiles catalogue shows that this estimate is reasonable: Seven out of ten stars within 4$\degree$ from the SN have been measured to be unpolarised, while the remaining three  exhibit low levels of polarisation, indicating that the Galactic ISP is less than $\sim 0.15$\% near the SN line-of-sight. The star also fully probes the Galactic dust column. It is found at a galactic latitude $b=-65.03$\degree~and parallax $p=0.56$\,mas corresponding position $\sim1600$\,pc above the Galactic plane.

\subsection{SN\,2019neq}
For SN\,2019neq there are two bright point sources in the NOT/ALFOSC FOV (see Figure \ref{fig:NOT_stamps}). The $e$ beam of Star 1 is found on the edge of the $o$ beam image and the polarisation measurements may not be reliable. Star 2 was saturated in the first epoch of observation, and the second epoch of observation is contaminated by the bright moon. Thus, we estimate the Galactic ISP based on the observations of Star 2 at epochs three and four. We find $Q_\mathrm{ISP}=0.27\pm0.03\%$ $U_\mathrm{ISP}= -0.18\pm0.03\%$, which is consistent with the polarisation of $P_\mathrm{ISP}= 0.18\pm0.20\%$ of an unpolarised star within 3$^{\circ}$ from the location of the SN. The star is found at $650$\,pc above the Galactic plane ($b=28.91$\degree~and $p=0.74$\,mas) and should probe the dust column sufficiently.

\subsection{SN\,2020tcw}
\begin{figure*}[h]
    \centering
     \begin{subfigure}[b]{0.33\textwidth}
         \centering
        \includegraphics[width=\textwidth]{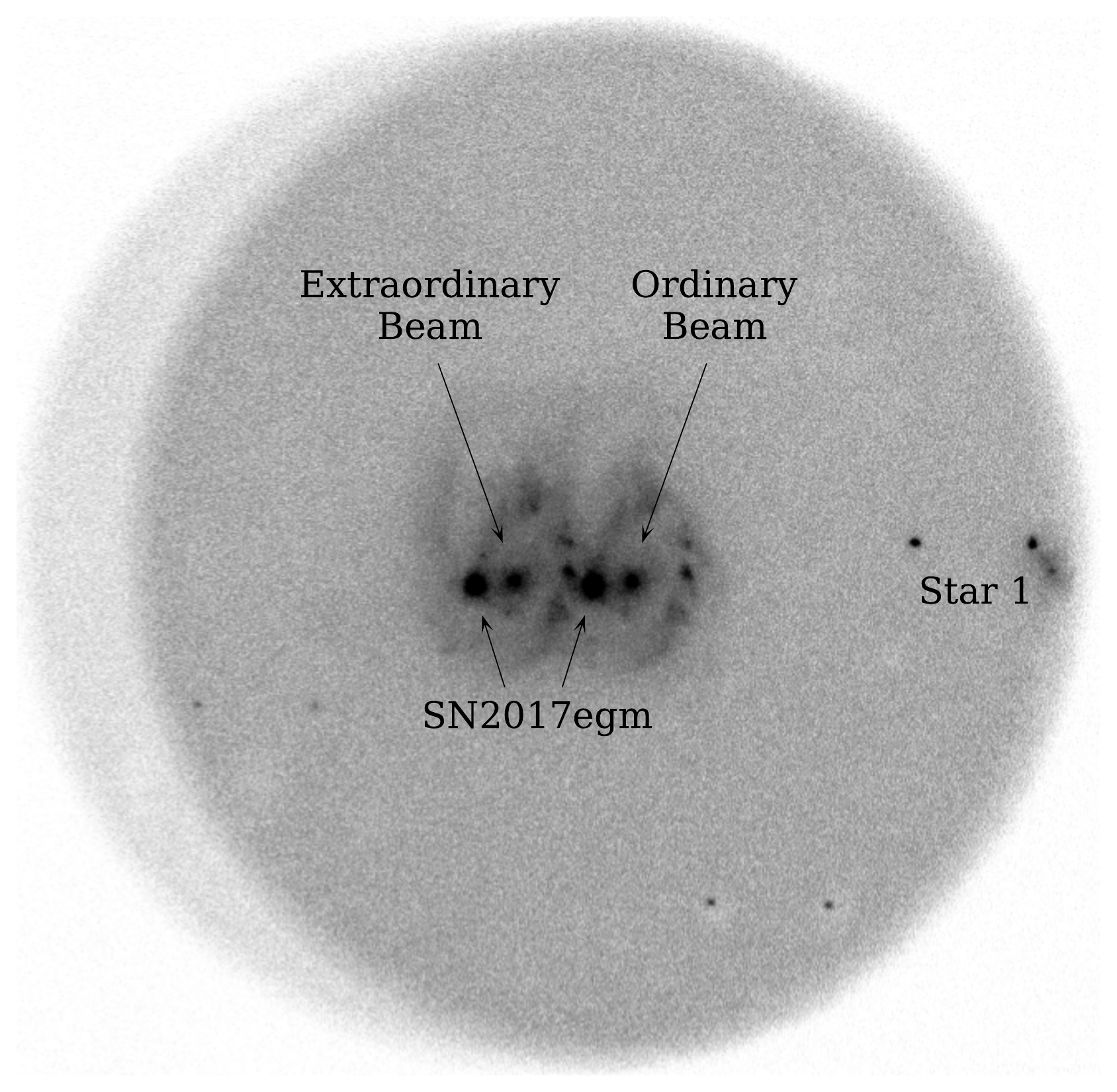}
    \end{subfigure} %
    \begin{subfigure}[b]{0.33\textwidth}
         \centering
        \includegraphics[width=\textwidth]{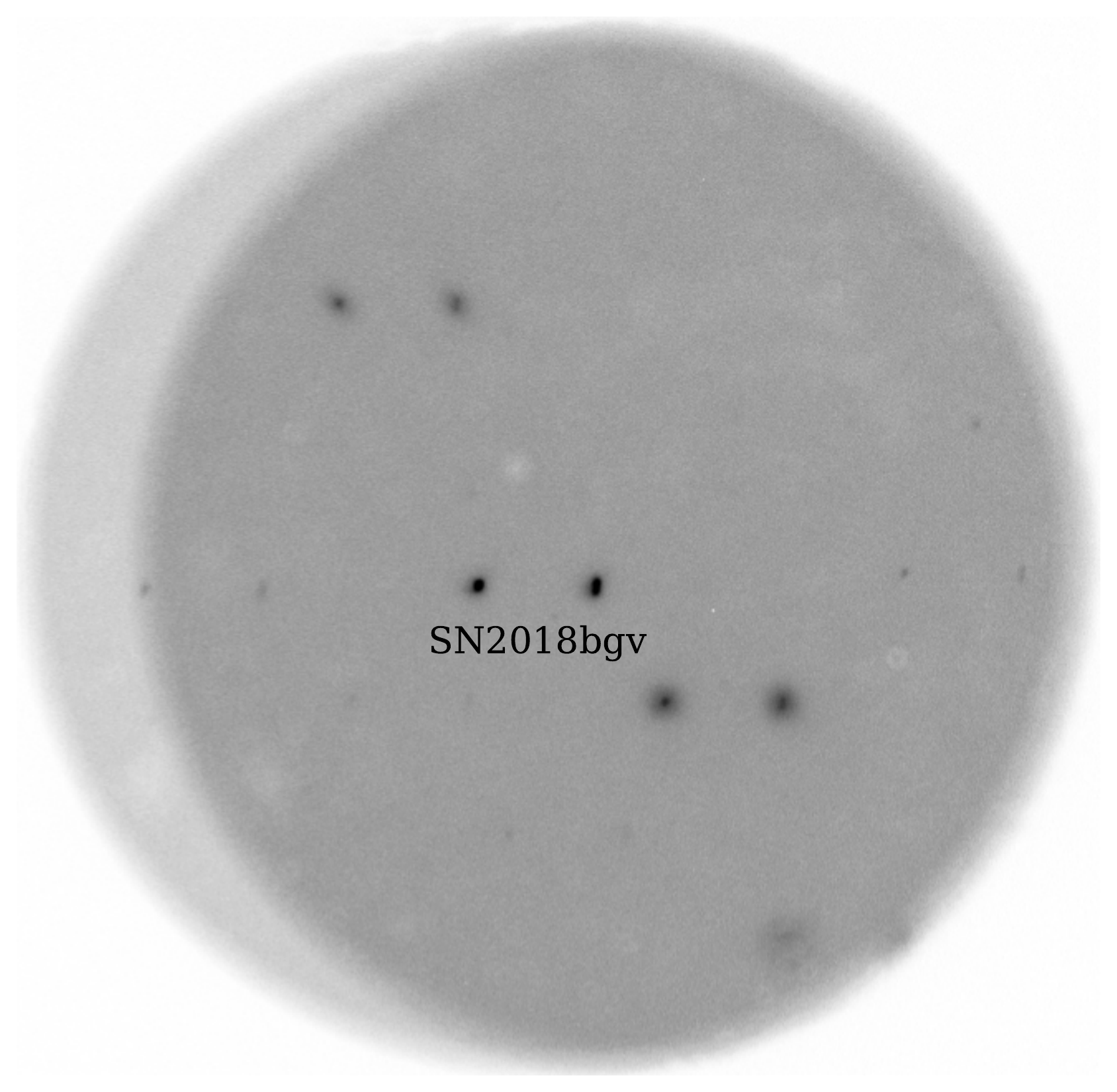}
    \end{subfigure} %
    \begin{subfigure}[b]{0.33\textwidth}
         \centering
        \includegraphics[width=\textwidth]{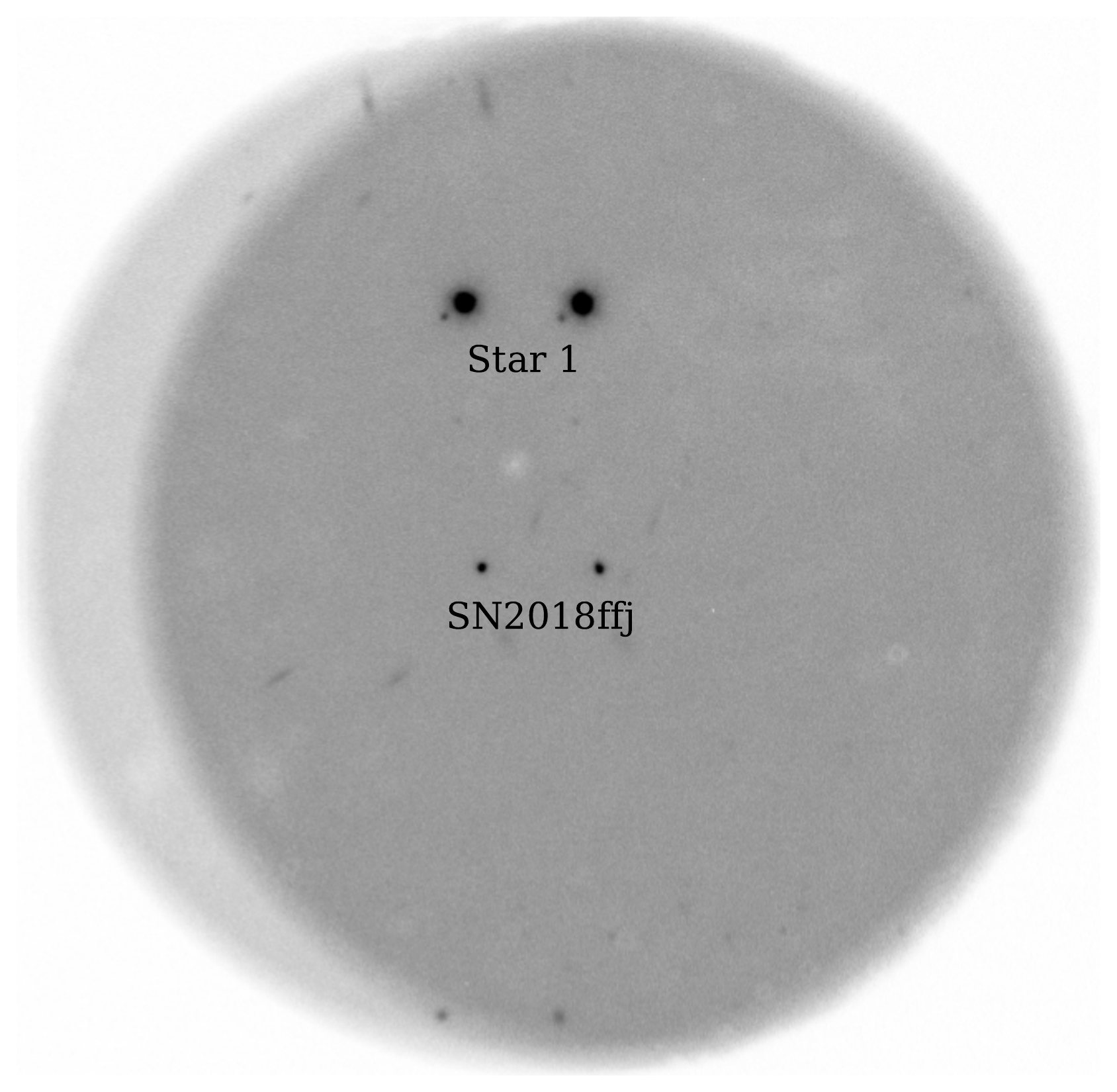}
    \end{subfigure} %
    \begin{subfigure}[b]{0.33\textwidth}
         \centering
        \includegraphics[width=\textwidth]{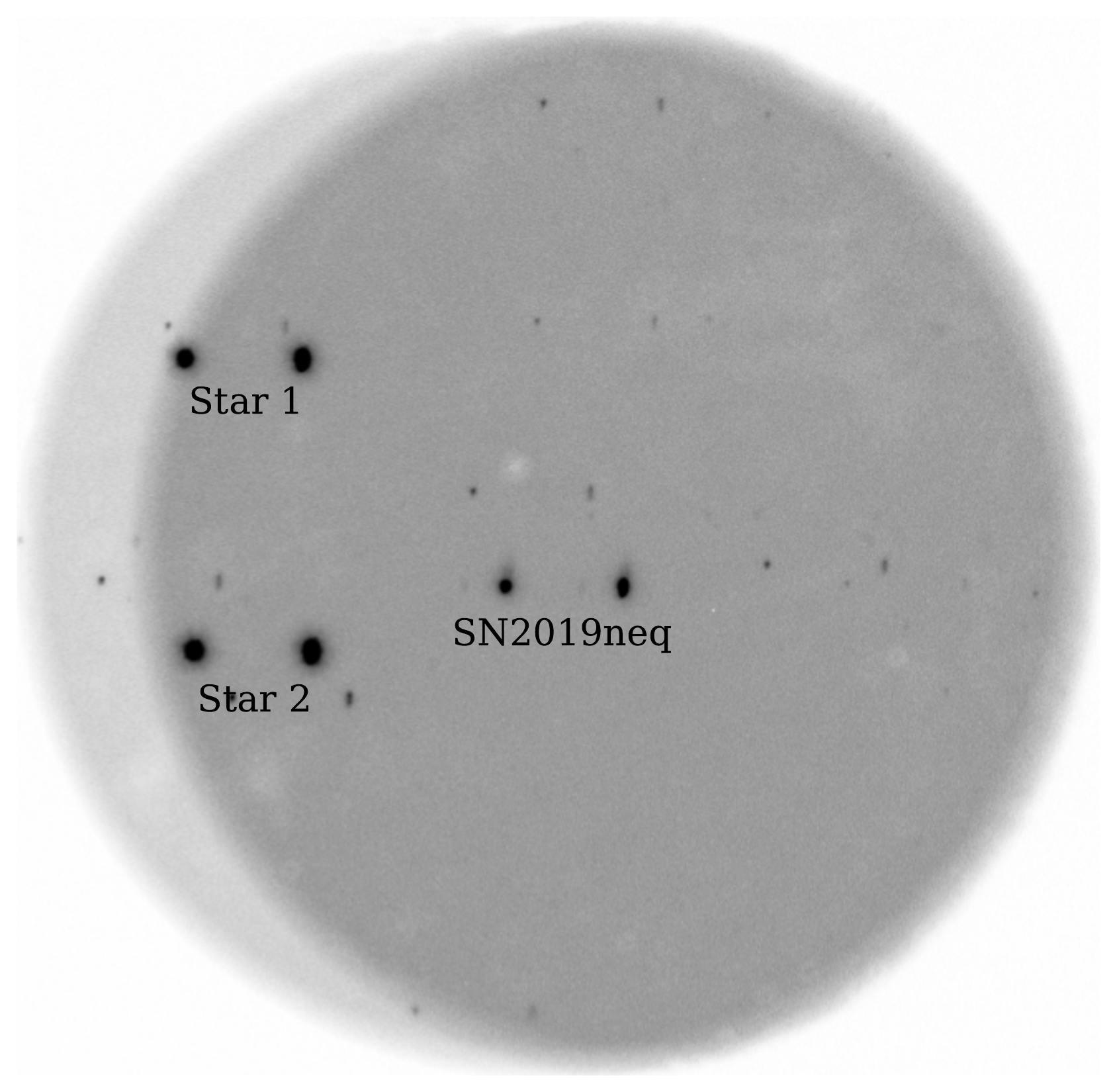}
    \end{subfigure} %
    \begin{subfigure}[b]{0.33\textwidth}
         \centering
        \includegraphics[width=\textwidth]{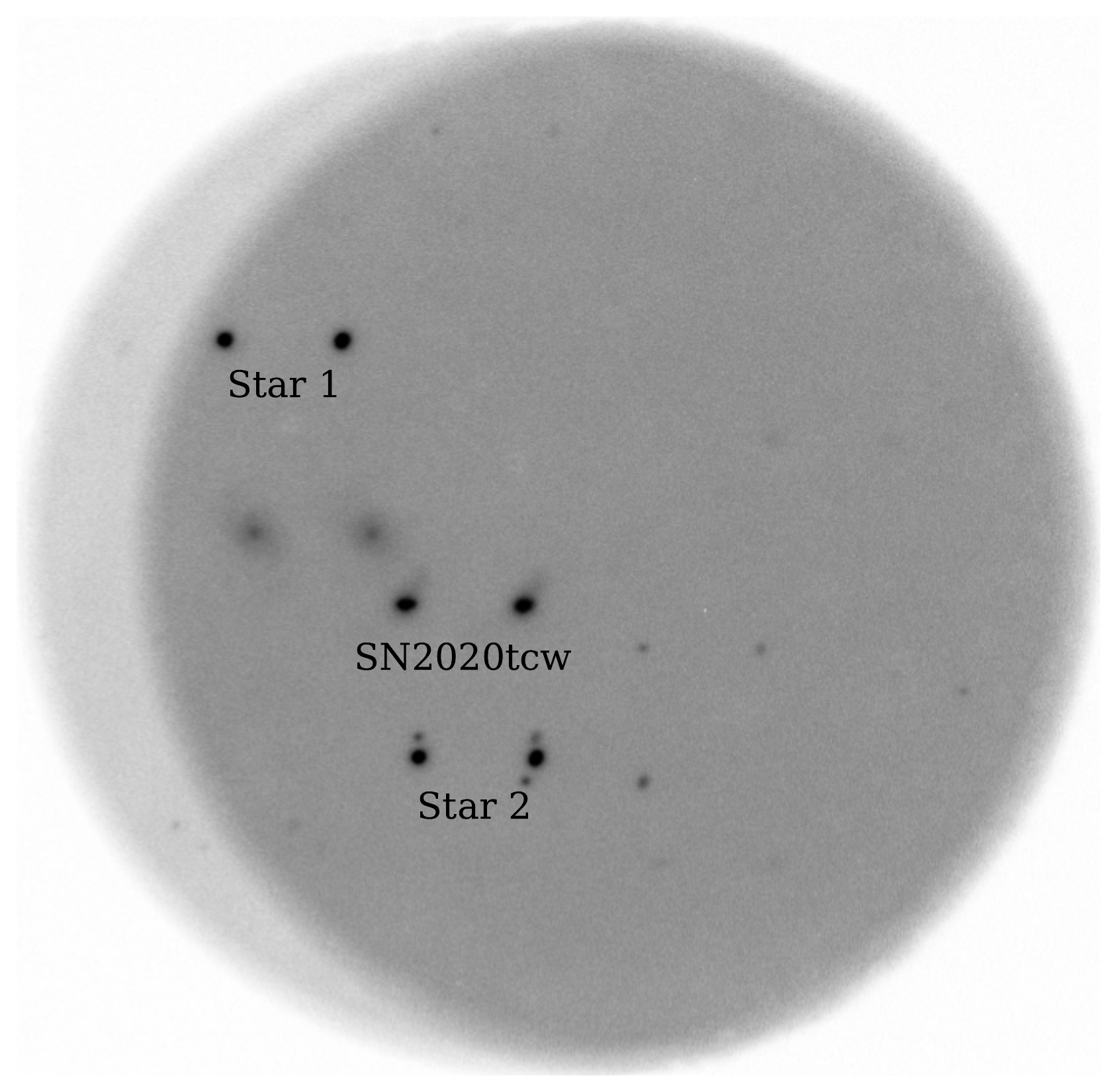}
    \end{subfigure} %

    \caption{Example NOT/ALFOSC polarimetry images for SN\,2017egm, SN\,2018bgv, SN\,2018ffj, SN\,2019neq and SN\,2020tcw. In the polarimetric mode, the $e$ (left) and $o$ beams (right) are overlaid with an offset of 15$\arcsec$. In addition to the SNe, bright point sources (i.e. stars) have been marked in the images. For SN\,2017egm, Star 1 is located on the edge of the $e$ beam image and is unreliable to estimate the ISP. No stars are present in the FOV of SN\,2018bgv. Star 1 was used estimate the ISP for SN\,2018ffj. For SN\,2019neq, only Star 2 is used to determine the ISP as Star 1 is found on the edge of the $o$ beam image resulting in an unreliable ISP estimate. Only Star 1 is used to estimate the ISP of SN\,2020tcw as Star 2 has several objects in the close vicinity and the ISP cannot be reliably identified.}
    \label{fig:NOT_stamps}
\end{figure*}

For SN\,2020tcw there are two bright stars in the ALFOSC FOV (see Figure \ref{fig:NOT_stamps}). Star 2 is in close vicinity of other fainter targets, which may contaminate the measurements. Therefore, we use Star 1 and estimate the Galactic ISP to be  $Q_\mathrm{ISP}=-0.16\pm0.11\%$ $U_\mathrm{ISP}=-0.03\pm0.11\%$. The low Galactic ISP estimated for SN\,2020tcw is consistent with the zero-level polarisation towards the unpolarised stars within 2$^{\circ}$\,--\,4$^{\circ}$ from the location of the SN. The star is also located at a distance of $480$\,pc from the Galactic plane (latitude $b=55.23$\degree~and $p=1.71$\,mas) and should provide a reliable estimate for the Galactic ISP.

\begin{table}
    \def\arraystretch{1.1}%
    \setlength\tabcolsep{5pt}
    \centering
    \fontsize{10}{12}\selectfont
    \caption{The estimated ISPs for the individual epochs of VLT/FORS2 imaging polarimetry of SN\,2017gci with the number of cycles per epoch ($n$). Note that second epoch is based on one cycle of observation making it less reliable than the other three and is thus excluded in estimating the final value.}
    \begin{threeparttable}
    
    \begin{tabular}{l c r r r}
    \hline
    \hline
    \multicolumn{1}{c}{Date} &  \multicolumn{1}{c}{$n$} & \multicolumn{1}{c}{$Q_\mathrm{ISP}$ (\%)}  & \multicolumn{1}{c}{$U_\mathrm{ISP}$ (\%)}   & \multicolumn{1}{c}{$P_\mathrm{ISP}$ (\%)}     \\
    \hline
2017-08-29 & 2 & $-0.54 \pm 0.02$ & $-0.01 \pm 0.02$ & $ 0.54 \pm 0.02$ \\
2017-09-23 & 1 & $-0.52 \pm 0.03$ & $-0.23 \pm 0.03$ & $ 0.57 \pm 0.03$ \\
2017-10-19 & 2 & $-0.48 \pm 0.02$ & $-0.05 \pm 0.02$ & $ 0.48 \pm 0.02$ \\
2017-12-17 & 3 & $-0.50 \pm 0.02$ & $-0.01 \pm 0.02$ & $ 0.50 \pm 0.02$ \\
    \hline
\textbf{Final} & - & $-0.50 \pm 0.01$ & $-0.02 \pm 0.01$ & $ 0.50 \pm 0.01$ \\
    \hline
    \hline
    \end{tabular}

    \end{threeparttable}    
\label{tab:isp_gci}
\end{table}

\section{Tables} \label{app:Tables}
\onecolumn
\begin{table*}[h]
	\centering
	   \def\arraystretch{0.96}%
    \setlength\tabcolsep{10pt}
    \centering
    \fontsize{10}{12}\selectfont
	\caption{Observations log for SN\,2017gci in the spectropolarimetry (PMOS) and broad-band polarimetry (IPOL) modes.}
	\begin{tabular}{cccccc} 
    \hline
	\hline
Date and time & Mode & Filter & HWP angle & Exposure & Seeing \\
(UTC) & &  & ($^\circ$) & (s) & ('')\\
\hline
2017-08-25 08:39:01 & PMOS & free & 0.0 & 750 & 0.77 \\
2017-08-25 08:52:25 & PMOS & free & 45.0 & 750 & 1.1 \\
2017-08-25 09:05:40 & PMOS & free & 22.5 & 750 & 1.21 \\
2017-08-25 09:19:04 & PMOS & free & 67.5 & 750 & 1.06 \\
2017-08-26 08:30:16 & PMOS & free & 0.0 & 750 & 0.75 \\
2017-08-26 08:43:40 & PMOS & free & 45.0 & 750 & 0.88 \\
2017-08-26 08:56:55 & PMOS & free & 22.5 & 750 & 0.98 \\
2017-08-26 09:10:18 & PMOS & free & 67.5 & 750 & 0.91 \\
 &  &  &  &  & \\
2017-08-29 09:31:51 & IPOL & $V_\mathrm{HIGH}$ & 0.0 & 80 & 0.55 \\
2017-08-29 09:33:53 & IPOL & $V_\mathrm{HIGH}$ & 45.0 & 80 & 0.5 \\
2017-08-29 09:35:49 & IPOL & $V_\mathrm{HIGH}$ & 22.5 & 80 & 0.47 \\
2017-08-29 09:37:52 & IPOL & $V_\mathrm{HIGH}$ & 67.5 & 80 & 0.47 \\
2017-08-29 09:46:32 & IPOL & $V_\mathrm{HIGH}$ & 0.0 & 80 & 0.5 \\
2017-08-29 09:48:36 & IPOL & $V_\mathrm{HIGH}$ & 45.0 & 80 & 0.53 \\
2017-08-29 09:50:31 & IPOL & $V_\mathrm{HIGH}$ & 22.5 & 80 & 0.51 \\
2017-08-29 09:52:34 & IPOL & $V_\mathrm{HIGH}$ & 67.5 & 80 & 0.52 \\
 &  &  &  &  & \\
2017-09-22 07:15:51 & PMOS & free & 0.0 & 900 & 0.94 \\
2017-09-22 07:31:44 & PMOS & free & 45.0 & 900 & 0.67 \\
2017-09-22 07:47:29 & PMOS & free & 22.5 & 900 & 0.7 \\
2017-09-22 08:03:23 & PMOS & free & 67.5 & 900 & 0.82 \\
2017-09-22 08:21:44 & PMOS & free & 0.0 & 900 & 0.86 \\
2017-09-22 08:37:38 & PMOS & free & 45.0 & 900 & 0.8 \\
2017-09-22 08:53:24 & PMOS & free & 22.5 & 900 & 0.77 \\
2017-09-22 09:09:18 & PMOS & free & 67.5 & 900 & 0.61 \\
 &  &  &  &  & \\
2017-09-23 08:16:55 & IPOL & $V_\mathrm{HIGH}$ & 0.0 & 140 & 0.69 \\
2017-09-23 08:19:58 & IPOL & $V_\mathrm{HIGH}$ & 45.0 & 140 & 0.65 \\
2017-09-23 08:22:53 & IPOL & $V_\mathrm{HIGH}$ & 22.5 & 140 & 0.68 \\
2017-09-23 08:25:56 & IPOL & $V_\mathrm{HIGH}$ & 67.5 & 140 & 0.68 \\
 &  &  &  &  & \\
2017-10-19 07:14:19 & IPOL & $V_\mathrm{HIGH}$ & 0.0 & 120 & 0.57 \\
2017-10-19 07:17:03 & IPOL & $V_\mathrm{HIGH}$ & 45.0 & 150 & 0.56 \\
2017-10-19 07:20:09 & IPOL & $V_\mathrm{HIGH}$ & 22.5 & 150 & 0.6 \\
2017-10-19 07:23:22 & IPOL & $V_\mathrm{HIGH}$ & 67.5 & 150 & 0.58 \\
2017-10-19 07:27:32 & IPOL & $V_\mathrm{HIGH}$ & 0.0 & 150 & 0.6 \\
2017-10-19 07:30:46 & IPOL & $V_\mathrm{HIGH}$ & 45.0 & 150 & 0.66 \\
2017-10-19 07:33:51 & IPOL & $V_\mathrm{HIGH}$ & 22.5 & 150 & 0.68 \\
2017-10-19 07:37:03 & IPOL & $V_\mathrm{HIGH}$ & 67.5 & 150 & 0.61 \\
 &  &  &  &  & \\
2017-12-17 03:34:35 & IPOL & $V_\mathrm{HIGH}$ & 0.0 & 400 & 0.72 \\
2017-12-17 03:41:58 & IPOL & $V_\mathrm{HIGH}$ & 45.0 & 400 & 0.74 \\
2017-12-17 03:49:13 & IPOL & $V_\mathrm{HIGH}$ & 22.5 & 400 & 0.87 \\
2017-12-17 03:56:35 & IPOL & $V_\mathrm{HIGH}$ & 67.5 & 400 & 0.73 \\
2017-12-17 04:04:41 & IPOL & $V_\mathrm{HIGH}$ & 0.0 & 400 & 0.8 \\
2017-12-17 04:12:03 & IPOL & $V_\mathrm{HIGH}$ & 45.0 & 400 & 0.74 \\
2017-12-17 04:19:18 & IPOL & $V_\mathrm{HIGH}$ & 22.5 & 400 & 0.58 \\
2017-12-17 04:26:41 & IPOL & $V_\mathrm{HIGH}$ & 67.5 & 400 & 0.52 \\
2017-12-17 04:34:42 & IPOL & $V_\mathrm{HIGH}$ & 0.0 & 400 & 0.52 \\
2017-12-17 04:42:04 & IPOL & $V_\mathrm{HIGH}$ & 45.0 & 400 & 0.51 \\
2017-12-17 04:49:20 & IPOL & $V_\mathrm{HIGH}$ & 22.5 & 400 & 0.51 \\
2017-12-17 04:56:43 & IPOL & $V_\mathrm{HIGH}$ & 67.5 & 400 & 0.52 \\
    \hline
		\hline
	\end{tabular}
\label{tab:gci_obs_log}
\end{table*}

\twocolumn
\begin{table}
	\centering
    \def\arraystretch{1.}%
    \setlength\tabcolsep{2pt}
    \centering
    \fontsize{10}{12}\selectfont
    \caption{Observations log for SN\,2017egm.}
	\begin{tabular}{cccc} 
\hline
\hline
Date and time  & Filter & HWP angle & Exposure \\
(UTC) & & ($^\circ$) & (s) \\
\hline
2017-06-17 22:00:05  & R   & $0.0    $ & $   40$ \\
2017-06-17 22:00:55  & R   & $22.5   $ & $   40$ \\
2017-06-17 22:01:44  & R   & $45.0   $ & $   40$ \\
2017-06-17 22:02:34  & R   & $67.5   $ & $   40$ \\
\\
2017-06-19 22:02:06  & V   & $0.0    $ & $   40$ \\
2017-06-19 22:02:56  & V   & $22.5   $ & $   40$ \\
2017-06-19 22:03:45  & V   & $45.0   $ & $   40$ \\
2017-06-19 22:04:34  & V   & $67.5   $ & $   40$ \\
2017-06-19 22:06:18  & R   & $0.0    $ & $   40$ \\
2017-06-19 22:07:08  & R   & $22.5   $ & $   40$ \\
2017-06-19 22:07:58  & R   & $45.0   $ & $   40$ \\
2017-06-19 22:08:47  & R   & $67.5   $ & $   40$ \\
2017-06-19 22:10:16  & I   & $0.0    $ & $   40$ \\
2017-06-19 22:11:06  & I   & $22.5   $ & $   40$ \\
2017-06-19 22:11:55  & I   & $45.0   $ & $   40$ \\
2017-06-19 22:12:44  & I   & $67.5   $ & $   40$ \\
\\
2017-06-28 22:40:41  & V   & $0.0    $ & $   40$ \\
2017-06-28 22:41:30  & V   & $22.5   $ & $   40$ \\
2017-06-28 22:42:20  & V   & $45.0   $ & $   40$ \\
2017-06-28 22:43:10  & V   & $67.5   $ & $   40$ \\
2017-06-28 22:44:29  & R   & $0.0    $ & $   40$ \\
2017-06-28 22:45:19  & R   & $22.5   $ & $   40$ \\
2017-06-28 22:46:07  & R   & $45.0   $ & $   40$ \\
2017-06-28 22:46:57  & R   & $67.5   $ & $   40$ \\
2017-06-28 22:48:18  & I   & $0.0    $ & $   40$ \\
2017-06-28 22:49:07  & I   & $22.5   $ & $   40$ \\
2017-06-28 22:49:59  & I   & $45.0   $ & $   40$ \\
2017-06-28 22:50:48  & I   & $67.5   $ & $   40$ \\
\\
2017-12-30 00:48:20  & V   & $0.0    $ & $  200$ \\
2017-12-30 00:51:52  & V   & $22.5   $ & $  200$ \\
2017-12-30 00:55:25  & V   & $45.0   $ & $  200$ \\
2017-12-30 00:58:58  & V   & $67.5   $ & $  200$ \\
2017-12-30 01:02:32  & V   & $0.0    $ & $  200$ \\
2017-12-30 01:06:05  & V   & $22.5   $ & $  200$ \\
2017-12-30 01:09:38  & V   & $45.0   $ & $  200$ \\
2017-12-30 01:13:10  & V   & $67.5   $ & $  200$ \\
\hline
\hline
	\end{tabular}
\label{tab:egm_obs_log}
\end{table}

\begin{table}
	\centering
    \def\arraystretch{1.}%
    \setlength\tabcolsep{2pt}
    \centering
    \fontsize{10}{12}\selectfont
    \caption{Observations log for SN\,2018bgv.}
	\begin{tabular}{cccc} 
\hline
\hline
Date and time  & Filter & HWP angle & Exposure \\
(UTC) & & ($^\circ$) & (s) \\
\hline
2018-05-24 00:43:09  & V   & $0.0    $ & $  500$ \\
2018-05-24 00:51:39  & V   & $22.5   $ & $  500$ \\
2018-05-24 01:00:11  & V   & $45.0   $ & $  500$ \\
2018-05-24 01:08:41  & V   & $67.5   $ & $  500$ \\
\\
2018-06-04 21:11:26  & V   & $0.0    $ & $  500$ \\
2018-06-04 21:19:56  & V   & $22.5   $ & $  500$ \\
2018-06-04 21:28:26  & V   & $45.0   $ & $  500$ \\
2018-06-04 21:36:55  & V   & $67.5   $ & $  500$ \\
\\
2018-06-12 22:20:44  & V   & $0.0    $ & $  600$ \\
2018-06-12 22:30:54  & V   & $22.5   $ & $  600$ \\
2018-06-12 22:41:04  & V   & $45.0   $ & $  600$ \\
2018-06-12 22:51:14  & V   & $67.5   $ & $  600$ \\
\\
2018-06-26 21:20:33  & V   & $0.0    $ & $  400$ \\
2018-06-26 21:27:23  & V   & $22.5   $ & $  400$ \\
2018-06-26 21:34:15  & V   & $45.0   $ & $  400$ \\
2018-06-26 21:41:05  & V   & $67.5   $ & $  400$ \\
2018-06-26 21:47:56  & V   & $0.0    $ & $  400$ \\
2018-06-26 21:54:46  & V   & $22.5   $ & $  400$ \\
2018-06-26 22:01:36  & V   & $45.0   $ & $  400$ \\
2018-06-26 22:08:26  & V   & $67.5   $ & $  400$ \\
\\
2018-07-12 21:25:25  & V   & $0.0    $ & $  550$ \\
2018-07-12 21:34:46  & V   & $22.5   $ & $  550$ \\
2018-07-12 21:44:07  & V   & $45.0   $ & $  550$ \\
2018-07-12 21:53:27  & V   & $67.5   $ & $  550$ \\
2018-07-12 22:02:48  & V   & $0.0    $ & $  550$ \\
2018-07-12 22:12:08  & V   & $22.5   $ & $  550$ \\
2018-07-12 22:21:28  & V   & $45.0   $ & $  550$ \\
2018-07-12 22:30:49  & V   & $67.5   $ & $  550$ \\
\hline
\hline
	\end{tabular}
\label{tab:bgv_obs_log}
\end{table}

\begin{table}
	\centering
    \def\arraystretch{1.}%
    \setlength\tabcolsep{2pt}
    \centering
    \fontsize{10}{12}\selectfont
    \caption{Observations log for SN\,2018ffj.}
	\begin{tabular}{cccc} 
\hline
\hline
Date and time  & Filter & HWP angle & Exposure \\
(UTC) & & ($^\circ$) & (s) \\
\hline
2018-08-26 04:46:16  & V   & $0.0    $ & $  600$ \\
2018-08-26 04:56:26  & V   & $22.5   $ & $  600$ \\
2018-08-26 05:06:38  & V   & $45.0   $ & $  600$ \\
2018-08-26 05:16:48  & V   & $67.5   $ & $  600$ \\
\\
2018-09-08 04:41:24  & V   & $0.0    $ & $  500$ \\
2018-09-08 04:49:54  & V   & $22.5   $ & $  500$ \\
2018-09-08 04:58:23  & V   & $45.0   $ & $  500$ \\
2018-09-08 05:06:53  & V   & $67.5   $ & $  500$ \\
2018-09-08 05:15:25  & V   & $0.0    $ & $  500$ \\
2018-09-08 05:23:55  & V   & $22.5   $ & $  500$ \\
2018-09-08 05:32:25  & V   & $45.0   $ & $  500$ \\
2018-09-08 05:40:55  & V   & $67.5   $ & $  500$ \\
\\
2018-09-19 02:49:14  & V   & $0.0    $ & $  500$ \\
2018-09-19 02:57:44  & V   & $22.5   $ & $  500$ \\
2018-09-19 03:06:14  & V   & $45.0   $ & $  500$ \\
2018-09-19 03:14:44  & V   & $67.5   $ & $  500$ \\
2018-09-19 03:23:16  & V   & $0.0    $ & $  500$ \\
2018-09-19 03:31:47  & V   & $22.5   $ & $  500$ \\
2018-09-19 03:40:17  & V   & $45.0   $ & $  500$ \\
2018-09-19 03:48:47  & V   & $67.5   $ & $  500$ \\
\hline
\hline
	\end{tabular}
\label{tab:ffj_obs_log}
\end{table}

\begin{table}
	\centering
    \def\arraystretch{1.}%
    \setlength\tabcolsep{2pt}
    \centering
    \fontsize{10}{12}\selectfont
    \caption{Observations log for SN\,2019neq.}
	\begin{tabular}{cccc} 
\hline
\hline
Date and time  & Filter & HWP angle & Exposure \\
(UTC) & & ($^\circ$) & (s) \\
\hline
2019-09-03 21:10:31  & V   & $0.0    $ & $  500$ \\
2019-09-03 21:18:59  & V   & $22.5   $ & $  500$ \\
2019-09-03 21:27:28  & V   & $45.0   $ & $  500$ \\
2019-09-03 21:35:56  & V   & $67.5   $ & $  500$ \\
\\
2019-09-08 21:11:49  & V   & $0.0    $ & $  500$ \\
2019-09-08 21:20:17  & V   & $22.5   $ & $  500$ \\
2019-09-08 21:28:46  & V   & $45.0   $ & $  500$ \\
2019-09-08 21:37:14  & V   & $67.5   $ & $  500$ \\
\\
2019-09-16 20:32:59  & V   & $0.0    $ & $  500$ \\
2019-09-16 20:41:27  & V   & $22.5   $ & $  500$ \\
2019-09-16 20:49:56  & V   & $45.0   $ & $  500$ \\
2019-09-16 20:58:24  & V   & $67.5   $ & $  500$ \\
\\
2019-09-25 21:13:24  & V   & $0.0    $ & $  650$ \\
2019-09-25 21:24:22  & V   & $22.5   $ & $  650$ \\
2019-09-25 21:35:20  & V   & $45.0   $ & $  650$ \\
2019-09-25 21:46:19  & V   & $67.5   $ & $  650$ \\
\hline
\hline
	\end{tabular}
\label{tab:neq_obs_log}
\end{table}

\begin{table}
	\centering
    \def\arraystretch{1.}%
    \setlength\tabcolsep{2pt}
    \centering
    \fontsize{10}{12}\selectfont
    \caption{Observations log for SN\,2020tcw.}
	\begin{tabular}{cccc} 
\hline
\hline
Date and time  & Filter & HWP angle & Exposure \\
(UTC) & & ($^\circ$) & (s) \\
\hline
2020-10-15 19:51:17  & V   & $0.0    $ & $  250$ \\
2020-10-15 19:55:36  & V   & $22.5   $ & $  250$ \\
2020-10-15 19:59:54  & V   & $45.0   $ & $  250$ \\
2020-10-15 20:04:12  & V   & $67.5   $ & $  250$ \\
\\
2021-01-19 05:59:18  & V   & $0.0    $ & $  500$ \\
2021-01-19 06:07:46  & V   & $22.5   $ & $  500$ \\
2021-01-19 06:16:15  & V   & $45.0   $ & $  500$ \\
2021-01-19 06:24:45  & V   & $67.5   $ & $  500$ \\
2021-01-19 06:33:15  & V   & $0.0    $ & $  500$ \\
2021-01-19 06:41:43  & V   & $22.5   $ & $  500$ \\
2021-01-19 06:50:12  & V   & $45.0   $ & $  500$ \\
2021-01-19 06:58:40  & V   & $67.5   $ & $  500$ \\
\hline
\hline
	\end{tabular}
\label{tab:tcw_obs_log}
\end{table}

\begin{table*}
    \def\arraystretch{1.1}%
    \setlength\tabcolsep{8pt}
    \centering
    \fontsize{10}{12}\selectfont
    \caption{Imaging polarimetry observations obtained with VLT/FORS2 for SN\,2017gci and NOT/ALFOSC for the five SLSNe-I. The mesurments of SN\,2018ibb are presented in Schulze et al. (in prep).}
    \begin{threeparttable}
    
    \begin{tabular}{r r r r r r r r r}
    \hline
    \hline
        \multicolumn{1}{c}{Date} &  \multicolumn{1}{c}{MJD} & \multicolumn{1}{c}{Phase\tnote{\bf a}} & \multicolumn{1}{c}{FWHM} & \multicolumn{1}{c}{S/N~Ratio} & \multicolumn{1}{c}{$Q$}    & \multicolumn{1}{c}{$U$}    & \multicolumn{1}{c}{$P$\tnote{\bf b}} & \multicolumn{1}{c}{$\theta$}     \\
             & \multicolumn{1}{c}{(d)}  & \multicolumn{1}{c}{(d)}   & \multicolumn{1}{c}{(pixel)} &     & \multicolumn{1}{c}{(\%)} & \multicolumn{1}{c}{(\%)} & \multicolumn{1}{c}{(\%)} & \multicolumn{1}{c}{($\degree$)}     \\
\hline
\multicolumn{9}{c}{\textbf{SN\,2017egm}} \\
\multicolumn{9}{c}{\textbf{V band}} \\
\hline
2017-06-19 & $57923.9$ &  $   -2.2$ & $   3.29$ & $ 601.80$ & $   0.34 \pm    0.36$ & $   0.15 \pm    0.40$ & $   0.25 \pm    0.38$ & $   11.9 \pm    28.3$ \\
2017-06-28 & $57932.9$ &  $    6.5$ & $   5.48$ & $ 501.13$ & $   0.23 \pm    0.37$ & $   0.02 \pm    0.40$ & $   0.13 \pm    0.39$ & $    2.5 \pm    45.6$ \\
\hline
\multicolumn{9}{c}{\textbf{R band}} \\
\hline
2017-06-17 & $57921.9$ &  $   -4.2$ & $   4.16$ & $ 568.31$ & $   0.15 \pm    0.36$ & $   0.15 \pm    0.40$ & $   0.12 \pm    0.38$ & $   22.5 \pm    51.4$ \\
2017-06-19 & $57923.9$ &  $   -2.2$ & $   3.81$ & $ 604.59$ & $   0.25 \pm    0.36$ & $   0.34 \pm    0.40$ & $   0.30 \pm    0.38$ & $   26.8 \pm    26.2$ \\
2017-06-28 & $57932.9$ &  $    6.5$ & $   5.84$ & $ 507.42$ & $   0.38 \pm    0.37$ & $  -0.07 \pm    0.40$ & $   0.26 \pm    0.39$ & $   -5.2 \pm    27.4$ \\
\hline
\multicolumn{9}{c}{\textbf{I band}} \\
\hline
2017-06-19 & $57923.9$ &  $   -2.2$ & $   3.10$ & $ 499.98$ & $  -0.16 \pm    0.37$ & $  -0.33 \pm    0.40$ & $   0.25 \pm    0.39$ & $  -57.9 \pm    31.1$ \\
2017-06-28 & $57932.9$ &  $    6.5$ & $   5.27$ & $ 394.91$ & $   0.49 \pm    0.38$ & $   0.44 \pm    0.42$ & $   0.54 \pm    0.40$ & $   21.0 \pm    17.4$ \\
\hline
\hline
\multicolumn{9}{c}{\textbf{SN\,2017gci}} \\
\multicolumn{9}{c}{\textbf{V band}} \\
\hline
2017-08-29 & $57994.4$ &  $    7.0$ & $   0.00$ & $1085.69$ & $  -0.05 \pm    0.06$ & $  -0.05 \pm    0.06$ & $   0.05 \pm    0.06$ & $  -67.6 \pm    25.5$ \\
2017-09-23 & $58019.3$ &  $   29.9$ & $   0.00$ & $ 767.70$ & $   0.06 \pm    0.09$ & $   0.20 \pm    0.09$ & $   0.19 \pm    0.09$ & $   36.1 \pm    12.5$ \\
2017-10-19 & $58045.3$ &  $   53.8$ & $   0.00$ & $ 542.84$ & $  -0.44 \pm    0.13$ & $   0.34 \pm    0.13$ & $   0.54 \pm    0.13$ & $   71.1 \pm     6.6$ \\
2017-12-17 & $58104.2$ &  $  108.0$ & $   0.00$ & $ 350.99$ & $  -0.32 \pm    0.20$ & $  -0.09 \pm    0.20$ & $   0.28 \pm    0.20$ & $  -82.0 \pm    17.4$ \\
\hline
\hline
\multicolumn{9}{c}{\textbf{SN\,2018bgv}} \\
\multicolumn{9}{c}{\textbf{V band}} \\
\hline
2018-05-24\tnote{\bf c} & $58262.0$ &  $    5.3$ & $   4.33$ & $ 279.36$ & $   0.14 \pm    0.25$ & $   0.03 \pm    0.26$ & $   0.08 \pm    0.26$ & $    6.0 \pm    50.1$ \\
2018-06-04 & $58273.9$ &  $   16.3$ & $   4.76$ & $ 354.19$ & $   0.49 \pm    0.21$ & $   0.62 \pm    0.19$ & $   0.76 \pm    0.20$ & $   25.8 \pm     7.2$ \\
2018-06-12 & $58281.9$ &  $   23.8$ & $   5.47$ & $ 269.97$ & $   0.18 \pm    0.26$ & $   0.37 \pm    0.26$ & $   0.34 \pm    0.26$ & $   32.0 \pm    18.1$ \\
2018-06-26\tnote{\bf d} & $58295.9$ &  $   36.7$ & $   5.99$ & $  89.45$ & $   3.69 \pm    0.81$ & $   1.75 \pm    0.78$ & $   4.01 \pm    0.80$ & $   12.7 \pm     5.7$ \\
2018-07-12\tnote{\bf d} & $58311.9$ &  $   51.6$ & $  10.84$ & $  82.91$ & $   3.27 \pm    0.86$ & $   0.75 \pm    0.84$ & $   3.25 \pm    0.85$ & $    6.5 \pm     7.3$ \\
\hline
\hline
\multicolumn{9}{c}{\textbf{SN\,2018ffj}} \\
\multicolumn{9}{c}{\textbf{V band}} \\
\hline
2018-08-26\tnote{\bf d} & $58356.2$ &  $   15.1$ & $   4.13$ & $  75.94$ & $   0.43 \pm    0.96$ & $  -1.70 \pm    0.91$ & $   1.50 \pm    0.94$ & $  -37.9 \pm    14.9$ \\
2018-09-08 & $58369.2$ &  $   25.6$ & $   3.33$ & $ 280.14$ & $  -0.44 \pm    0.25$ & $  -0.07 \pm    0.26$ & $   0.37 \pm    0.26$ & $  -85.2 \pm    16.1$ \\
2018-09-19 & $58380.2$ &  $   34.5$ & $   3.39$ & $ 176.82$ & $   0.37 \pm    0.40$ & $  -0.04 \pm    0.40$ & $   0.25 \pm    0.40$ & $   -3.5 \pm    30.8$ \\
\hline
\hline
\multicolumn{9}{c}{\textbf{SN\,2019neq}} \\
\multicolumn{9}{c}{\textbf{V band}} \\
\hline
2019-09-03 & $58729.9$ &  $   -3.7$ & $   3.69$ & $ 634.85$ & $   0.07 \pm    0.11$ & $  -0.12 \pm    0.11$ & $   0.11 \pm    0.11$ & $  -30.2 \pm    22.7$ \\
2019-09-08\tnote{\bf c} & $58734.9$ &  $    0.8$ & $   9.92$ & $ 237.53$ & $  -0.78 \pm    0.30$ & $   0.71 \pm    0.30$ & $   1.01 \pm    0.30$ & $   68.9 \pm     8.2$ \\
2019-09-16 & $58742.9$ &  $    8.0$ & $   3.52$ & $ 453.49$ & $  -0.32 \pm    0.16$ & $  -0.10 \pm    0.16$ & $   0.29 \pm    0.16$ & $  -81.7 \pm    14.0$ \\
2019-09-25 & $58751.9$ &  $   16.2$ & $   5.92$ & $ 434.88$ & $  -0.31 \pm    0.16$ & $   0.09 \pm    0.16$ & $   0.28 \pm    0.16$ & $   81.5 \pm    14.5$ \\
\hline
\hline
\multicolumn{9}{c}{\textbf{SN\,2020tcw}} \\
\multicolumn{9}{c}{\textbf{V band}} \\
\hline
2020-10-15 & $59137.8$ &  $    7.3$ & $   5.67$ & $ 576.84$ & $   0.15 \pm    0.17$ & $   0.01 \pm    0.17$ & $   0.10 \pm    0.17$ & $    1.1 \pm    30.9$ \\
2021-01-19 & $59233.3$ &  $   96.9$ & $   4.01$ & $ 502.44$ & $   0.03 \pm    0.18$ & $  -0.08 \pm    0.18$ & $   0.05 \pm    0.18$ & $  -34.0 \pm    57.0$ \\
\hline
\hline
    \end{tabular}
    \begin{tablenotes}
    \item[a] Rest frame days with respect to peak MJDs shown in Table \ref{tab:SLSN_list}.
    \item[b] Bias-corrected polarisation degree. Measurements whose error is larger than the value are consistent with zero polarisation.
    \item[c] The results are unreliable due to the high lunar illumination and are thus excluded.
    \item[d] The low S/N ratio makes the results unreliable and are thus excluded.
    \end{tablenotes}
    \end{threeparttable}    
\label{tab:impol_results}
\end{table*}

\clearpage
\onecolumn

\section{Figures}

\begin{figure*}[h]
    \centering
    \includegraphics[width=0.98\textwidth]{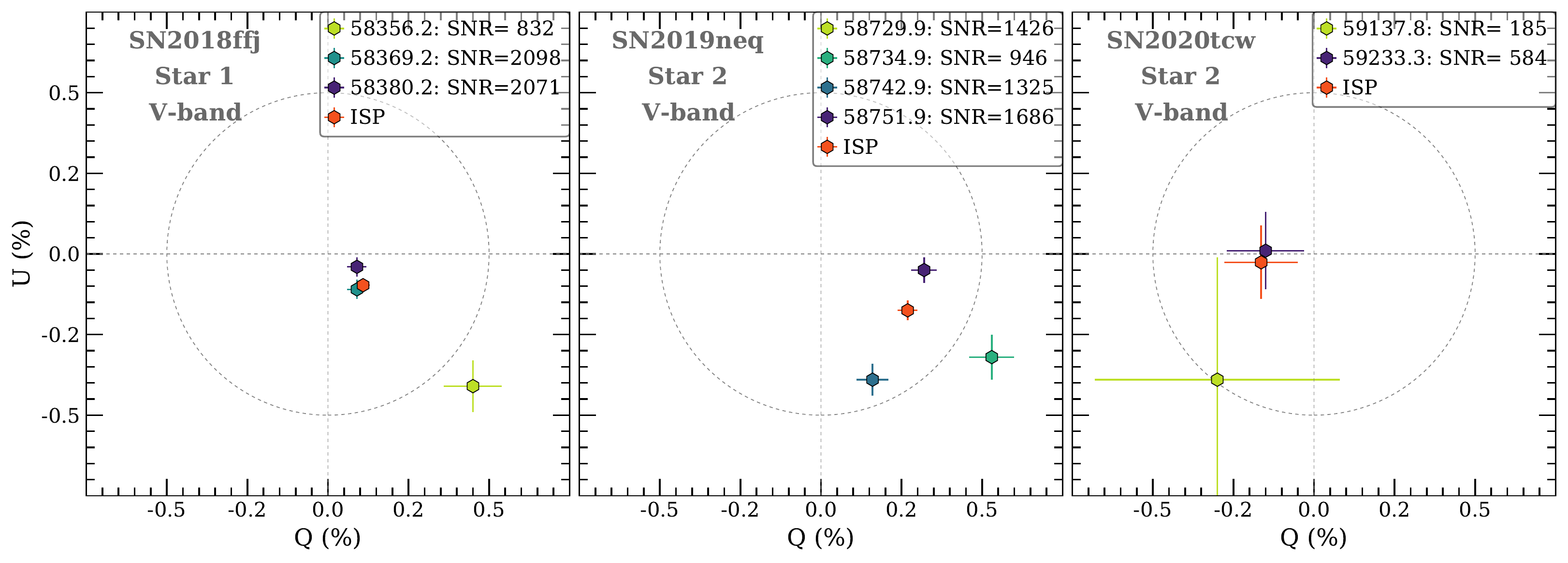}
    \caption{The $Q$\,--\,$U$ planes for the stars used to measure the ISP for SN\,2018ffj, SN\,2019neq and SN\,2020tcw. Note that the star was saturated in the first epoch of SN\,2019neq and it is excluded and that the first epoch of SN\,2018ffj and the second epoch of SN\,2019neq were excluded due to the high lunar illumination (see Section \ref{subsec:NOT_pola}).}
    \label{fig:NOT_QU_stars}
\end{figure*}

\begin{figure*}[h]
    \centering
    \includegraphics[width=0.98\textwidth]{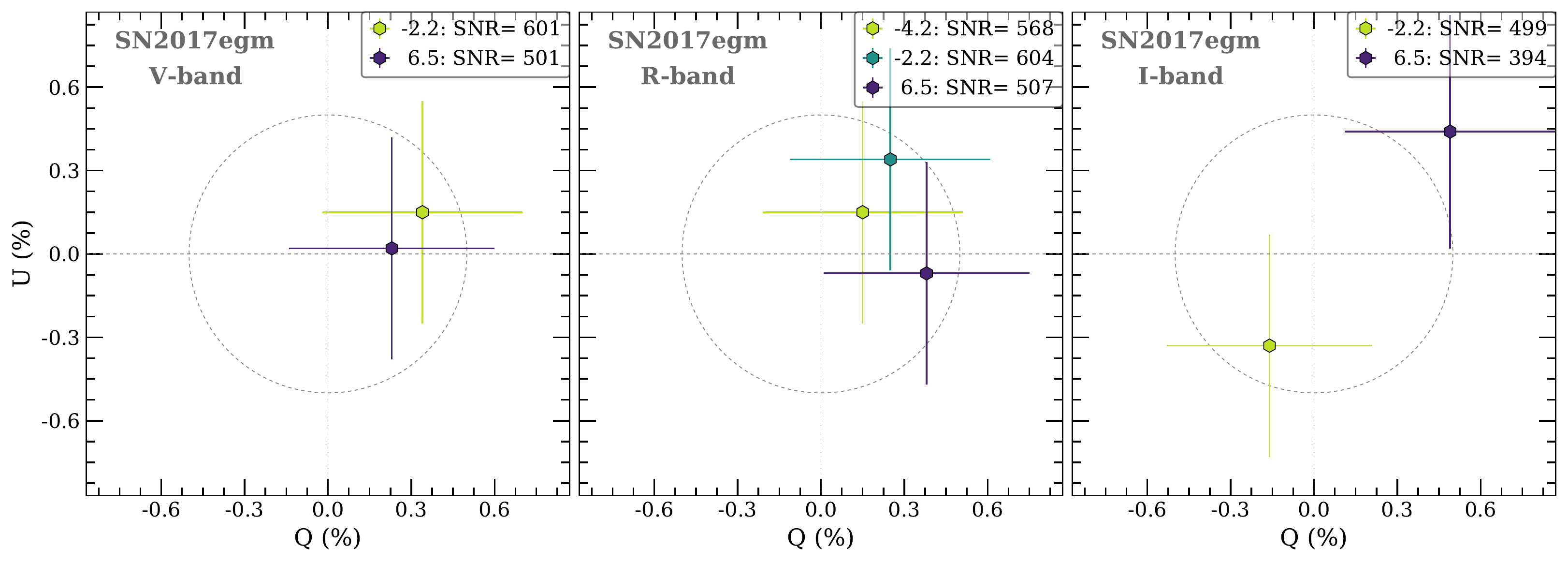}
    \caption{The $Q$\,--\,$U$ planes for the ISP corrected polarimetry of SN\,2017egm in $VRI$ bands. Note that the last epoch of NOT polarimetry is not shown in the figure due to extremely poor observing conditions (see Section \ref{subsec:NOT_pola})}
    \label{fig:QU_egm}
\end{figure*}

\begin{figure*}[h]
    \centering
    \begin{subfigure}[b]{0.341\textwidth}
         \centering
        \includegraphics[width=\textwidth]{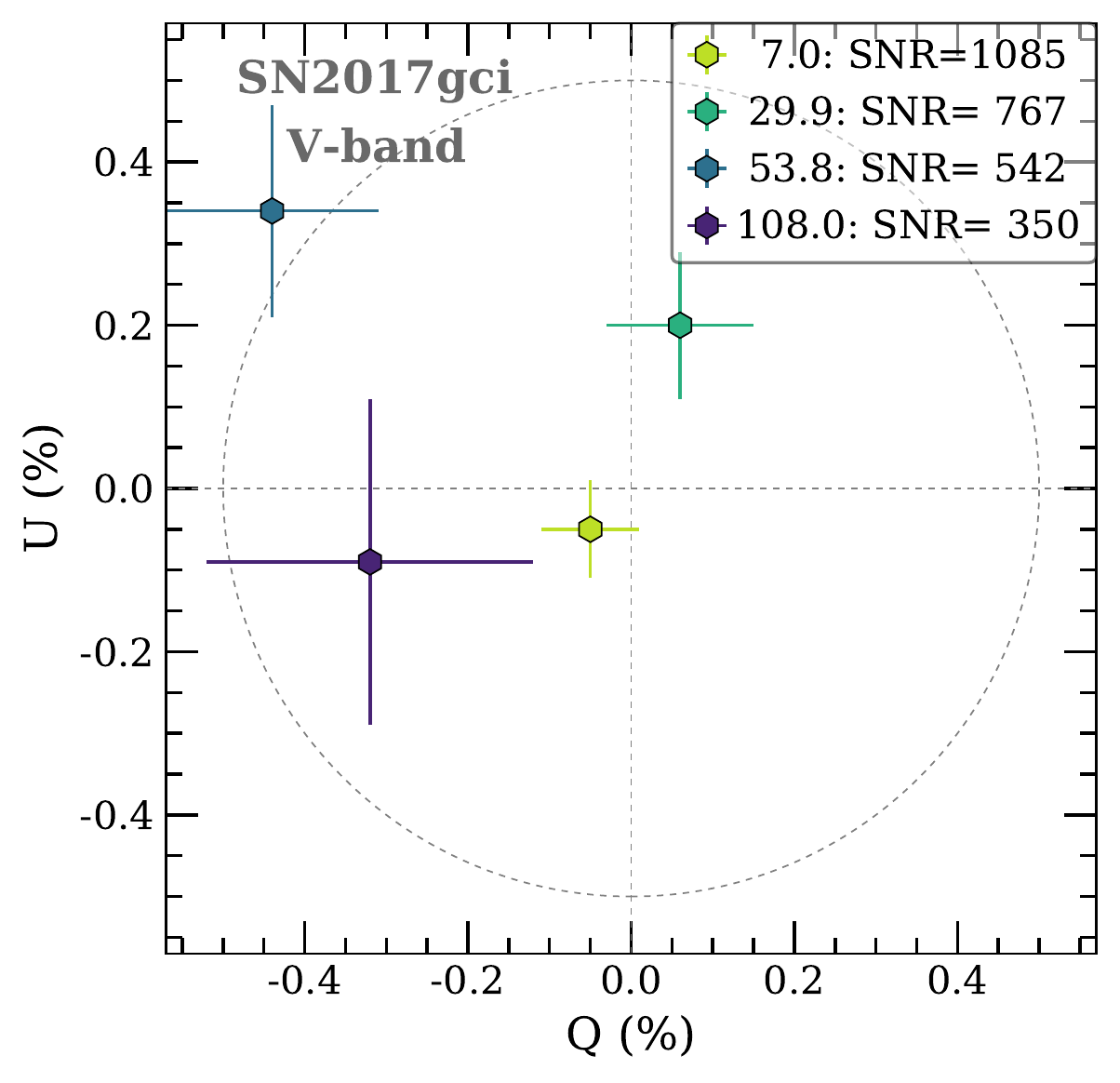}
    \end{subfigure} %
    \begin{subfigure}[b]{0.3245\textwidth}
         \centering
        \includegraphics[width=\textwidth]{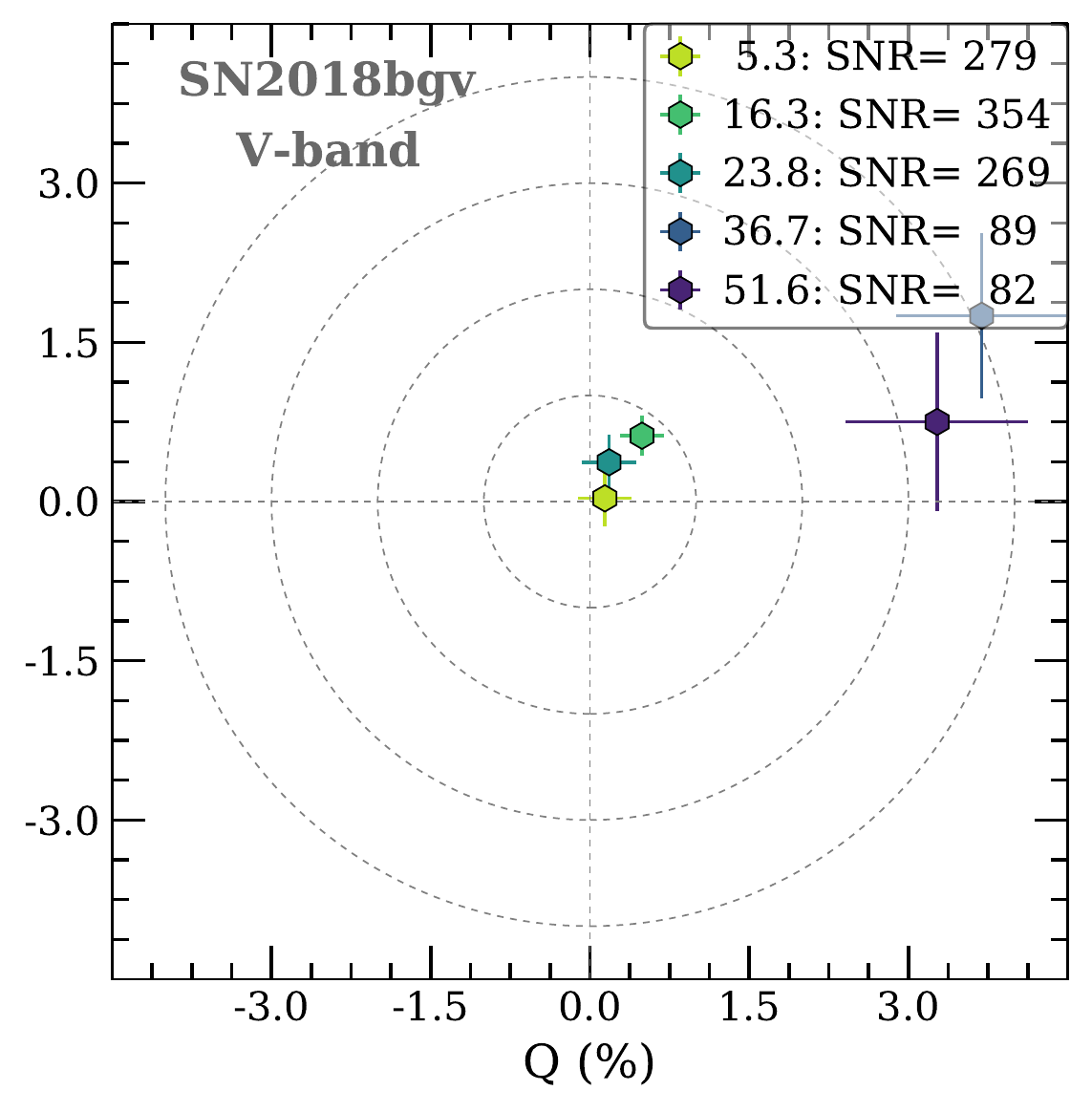}
    \end{subfigure} %
    \begin{subfigure}[b]{0.3245\textwidth}
         \centering
        \includegraphics[width=\textwidth]{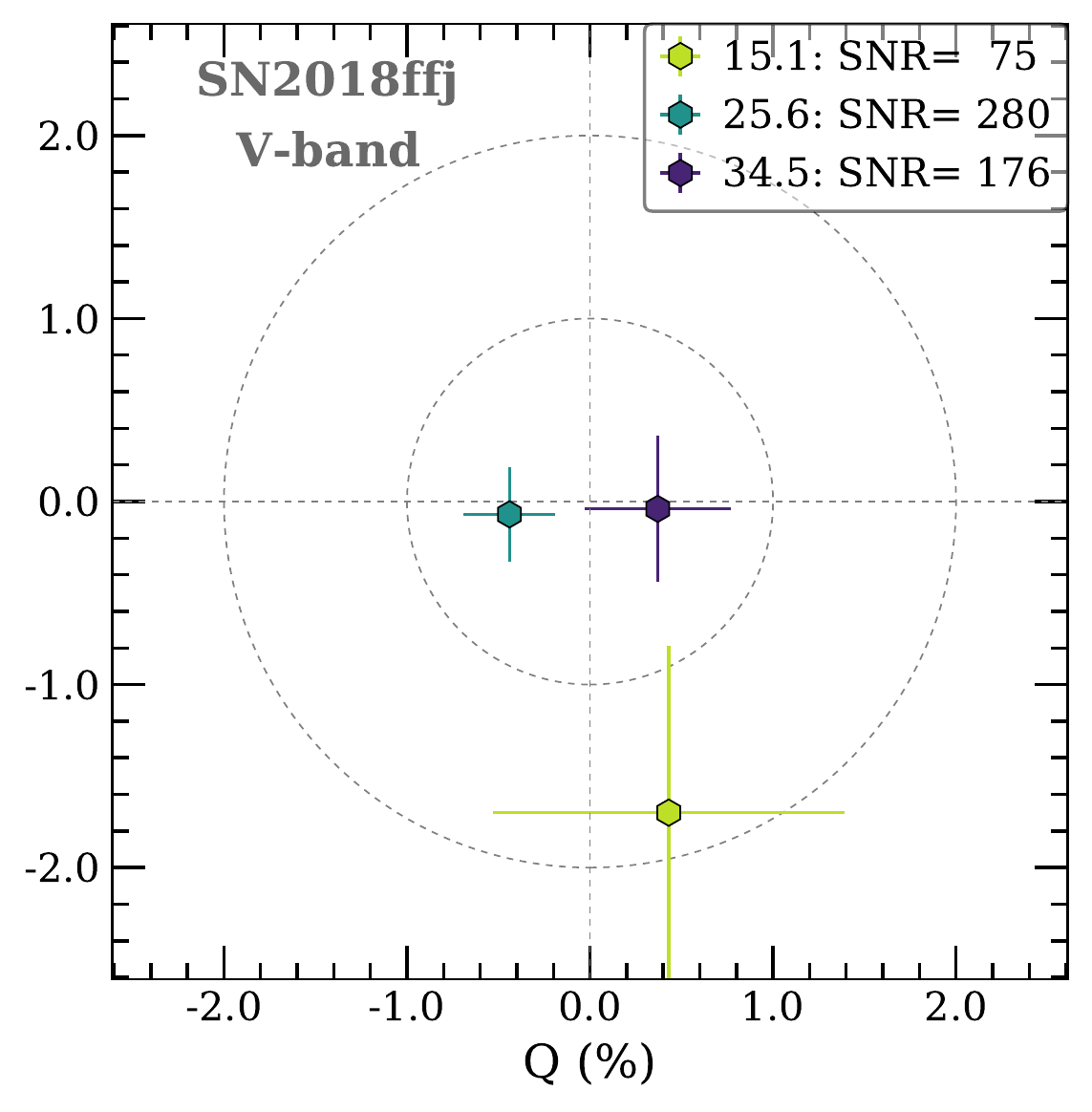}
    \end{subfigure} %
    \begin{subfigure}[b]{0.341\textwidth}
         \centering
        \includegraphics[width=\textwidth]{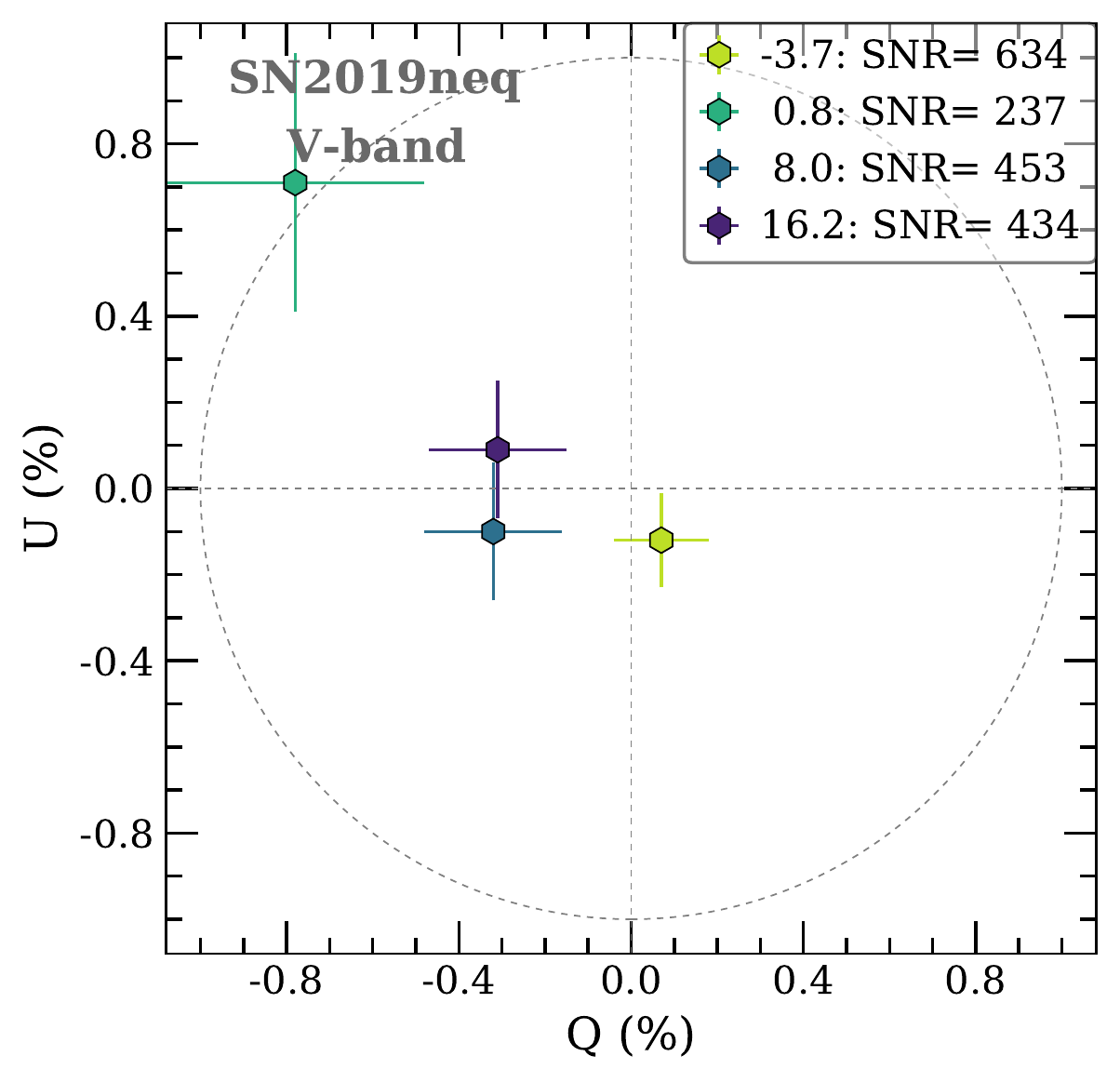}
    \end{subfigure} %
    \begin{subfigure}[b]{0.3245\textwidth}
         \centering
        \includegraphics[width=\textwidth]{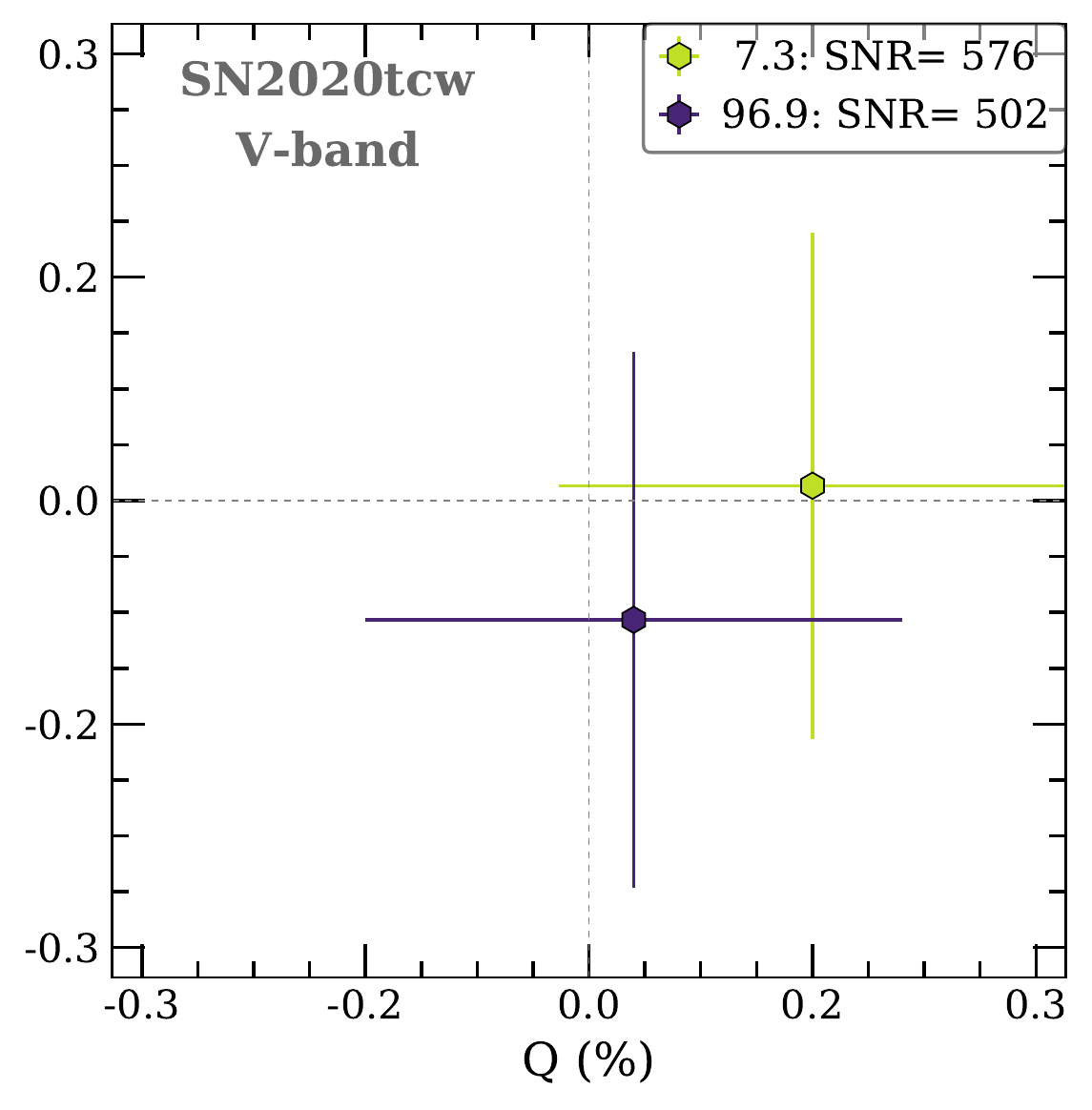}
    \end{subfigure} %
    \caption{The $Q$ -- $U$ planes of ISP-corrected polarimetry for SN\,2017gci, SN\,2018bgv, SN\,2018ffj, SN\,2019neq and SN\,2020tcw. Note that the two last epoch of SN\,2018bgv and first epoch of SN\,21018ffj were excluded from the analysis due to low S/N ratio and the second epoch of SN\,2019neq due to the bright Moon (see Section \ref{subsec:NOT_pola})}
    \label{fig:QU_SNe}
\end{figure*}

\begin{figure*}
    \centering
    \includegraphics[width=0.98\textwidth]{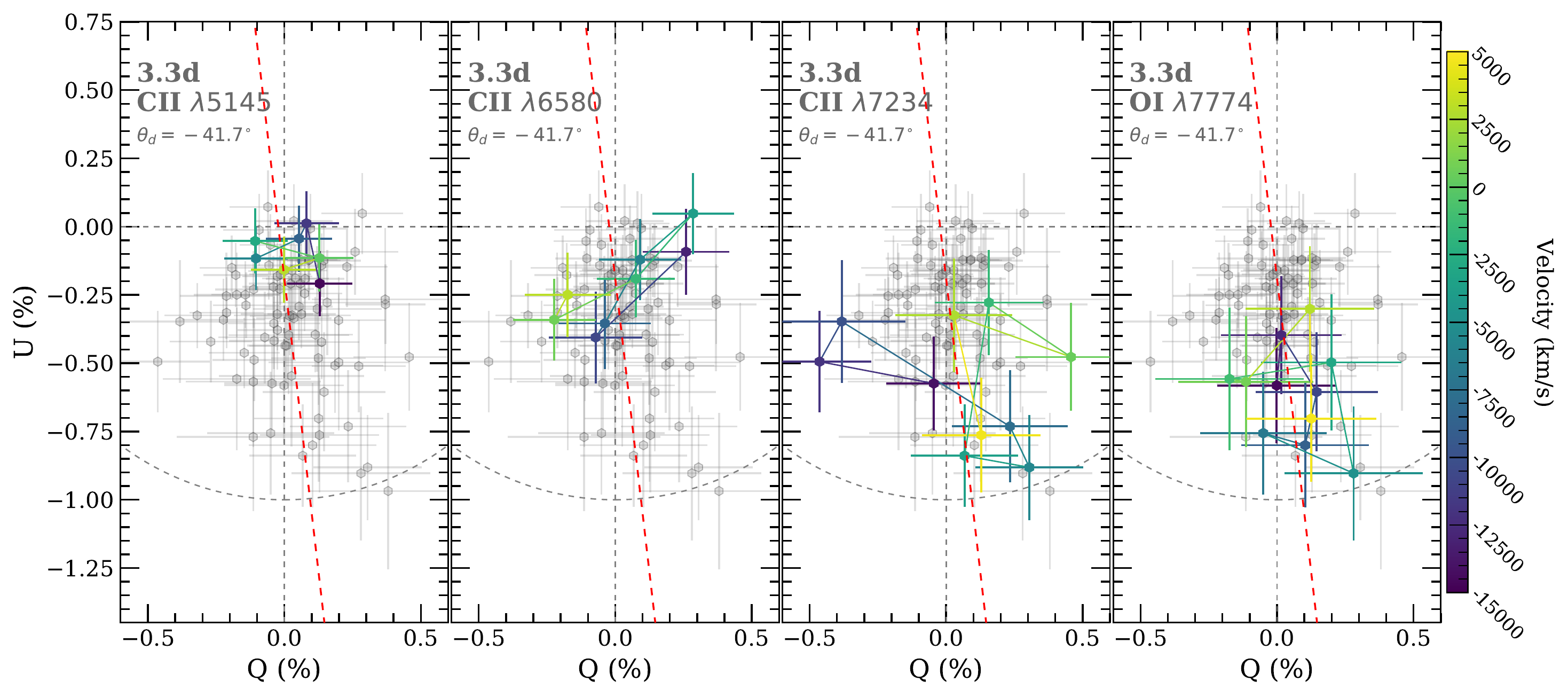}
    \caption{Notable line features on the $Q$\,--\,$U$ plane of SN\,2017gci at 3.3\,d. The line regions are highlighted by velocities as indicated by the colourbar, while other data points are shown in grey. The dominant axis from Figure \ref{fig:QU_plane_specpol_gci} is shown in dashed red. No loop-like structures are seen, but \ion{C}{II} $\lambda6580$ shows tentative deviation from the dominant direction.}
    \label{fig:QU_plane_loops_3d}
\end{figure*}

\begin{figure*}
    \centering
    \includegraphics[width=0.98\textwidth]{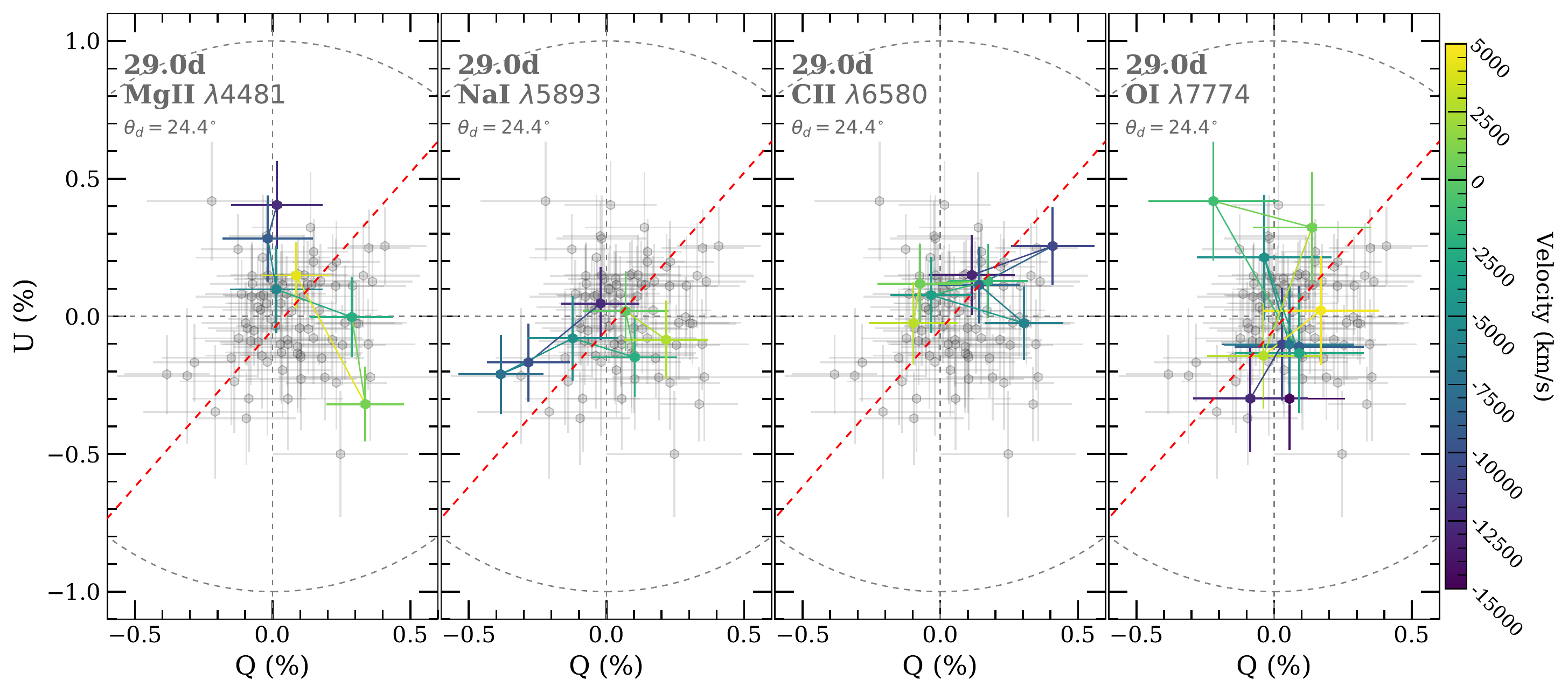}
    \caption{Same as Figure \ref{fig:QU_plane_loops_29d} but for $29.0\,$d. No notable loops are present. Despite no clear tendency was seen in Figure \ref{fig:QU_plane_specpol_gci}, \ion{Mg}{II} $\lambda4481$ is tentatively orthogonal to the shown dominant axis while \ion{Na}{I} $\lambda5893$ and possibly \ion{C}{II} $\lambda6580$ are along it.}
    \label{fig:QU_plane_loops_29d}
\end{figure*}

\end{appendix}

\end{document}